\documentclass[a4paper,12pt]{article}
\usepackage{soul}
\usepackage[usenames,dvipsnames]{color}
\usepackage{jheppub}
\usepackage{amsmath, amssymb, slashed, epsf, color, graphicx, latexsym}
\usepackage{epsfig}
\usepackage{graphics}
\newcommand{\CP}{{\cal P}}

\newcommand{\mK}{\mathcal{K}}

\newcommand{\mS}{\mathcal{S}}

\newcommand{\p}{\partial}

\begin{document}

\preprint{TIFR/TH/16-25}
\title{The large D black hole Membrane Paradigm at first subleading order}
\author[a]{Yogesh Dandekar,} 
\author[b] {Anandita De,}
\author[a]{Subhajit Mazumdar,}
\author[a]{Shiraz Minwalla}
\author[a]{and Arunabha Saha}
\affiliation[b]{Indian Institute of Science Education and Research Pune, Pune, India-411008}
\affiliation[a]{Tata Institute of Fundamental Research, Mumbai, India-400005}
\emailAdd{yogesh@theory.tifr.res.in}
\emailAdd{anu.anandita@gmail.com}
\emailAdd{subhajitmazumdar@theory.tifr.res.in}
\emailAdd{minwalla@theory.tifr.res.in}
\emailAdd{arunabha@theory.tifr.res.in}

\abstract{In the large D limit, and under certain circumstances, it has recently been demonstrated that 
black hole dynamics in asymptotically flat spacetime reduces to the dynamics of a non gravitational membrane propagating in flat $D$ dimensional spacetime. We demonstrate that this correspondence extends to all orders
in a $1/D$ expansion and outline a systematic method for deriving the corrected membrane equation 
in a power series expansion in $1/D$. As an illustration of our method we determine the first subleading
corrections to the membrane equations of motion. A qualitatively new effect at this order is that the divergence of the membrane velocity is nonzero and proportional to the square of the shear tensor reminiscent of 
the entropy current of hydrodynamics. As a test, we use our modified membrane 
equations to compute the corrections to frequencies of light quasinormal modes about the 
Schwarzschild black hole and find a perfect match with earlier computations performed directly in the
gravitational bulk.}

\maketitle

\section{Introduction}

It has recently been noted that the classical dynamics of black holes 
simplifies in the limit of a large number of dimensions. The key observation 
- first made by Emparan, Suzuki, Tanabe and collaborators in \cite{Emparan:2013moa,Emparan:2013xia,Emparan:2013oza,Emparan:2014cia,Emparan:2014jca,Emparan:2014aba,Emparan:2015rva} - is that black holes
at large $D$ have two effective length scales. The first of these, $r_0$,  
is the size of the black holes. The second 
is the thickness of the black hole's gravitational tail, i.e. the 
distance beyond the black hole event horizon  after which the gravitational 
potential rapidly decays to zero. In four dimensions 
the black hole size and thickness are comparable. In the large $D$ limit, 
however, the thickness of the gravitational tail turns out to  
scale like $r_0/D$ \cite{Emparan:2013moa} and so is much smaller than the the black hole 
size. 

This observation suggests the possibility of an effective `dimensional 
reduction' of black hole dynamics to the membrane region; a slab of spacetime
of thickness $1/D$ centered around the codimension one event horizon. 
In work done over the last year, this expectation has been borne out in
various contexts. In this paper we will focus on black holes propagating
in an otherwise unperturbed flat space. Assuming that $r_0$ (see above) 
and the length scale of variation along the horizon are both of order 
unity, the dimensional reduction described above was worked out to leading 
nontrivial order in the $1/D$ expansion for the most general nonlinear 
dynamical context in \cite{Bhattacharyya:2015dva,Bhattacharyya:2015fdk};
the special case of stationary solutions and their small fluctuations has 
also been studied at higher orders in the $1/D$ expansion in 
\cite{Emparan:2015hwa,Suzuki:2015iha,Tanabe:2015isb, Tanabe:2016opw}. 
In addition the dimensional reduction of small horizon ripples at 
length scale $1/\sqrt{D}$ about particular solutions (black strings or black 
branes in flat, $AdS$ or $dS$ space) has been studied in
\cite{Emparan:2015gva,Suzuki:2015axa,Tanabe:2015hda,Emparan:2016sjk,Tanabe:2016pjr}. Further developments were presented in 
\cite{Sadhu:2016ynd,Herzog:2016hob,Rozali:2016yhw,Chen:2015fuf,Giribet:2013wia,Prester:2013gxa,Chen:2016fuy}. 

In this paper we further develop the general nonlinear dynamical 
construction of \cite{Bhattacharyya:2015dva,Bhattacharyya:2015fdk}. 
In particular we demonstrate that the reduction of black hole dynamics to 
membrane dynamics, worked out to leading nontrivial order in the $1/D$ 
expansion in \cite{Bhattacharyya:2015dva,Bhattacharyya:2015fdk}, can be 
systematically generalized to every order in $1/D$. As an application 
of this systematic framework we explicitly work out the first subleading 
corrections to the membrane equations of motion in the $1/D$ expansion, and 
also determine the spacetimes dual to any particular membrane solution 
at next subleading order in the $1/D$ expansion. In this introduction 
we first review the leading order construction presented in 
\cite{Bhattacharyya:2015dva,Bhattacharyya:2015fdk} and then present our 
explicit higher order results. 

\subsection{Review of earlier work}

Consider a class of  $D$ dimensional metrics of the form  
\begin{equation}\label{ansatz0}
g_{MN}= \eta_{MN}+ \frac{ (n_M-u_M)(n_N-u_N) }{\psi^{D-3}} 
\end{equation}
The metrics \eqref{ansatz0} are parameterized by a smooth $D$ dimensional function $\psi$ and a smooth 
oneform field $u_M$. $n_M$ in \eqref{ansatz0} is the normal field to 
surfaces of constant $\psi$, (i.e. $n_M= \frac{\partial_M \psi} {\sqrt{\partial_P \psi \partial_Q\psi \eta^{PQ}}}$). The oneform field $u_M$ is assumed to be unit normalized (i.e. $u_N u_M \eta^{MN}=-1$) and tangent to surfaces of 
constant $\psi$ (i.e. $u_M n_N \eta^{MN}=0$). 

In order to gain intuition for spacetimes of the form \eqref{ansatz0} it is useful to first consider 
a special case. Working with coordinates in which the metric on Minkowski space takes the form 
$$ds^2=-dt^2 + dr^2 + r^2 d \Omega_{D-2}^2, $$ 
the choice $u=-dt$ and $ \psi= \frac{r}{r_0}$ 
turns \eqref{ansatz0} into the  metric of a Schwarzschild black hole of radius $r_0$ 
in the so called Kerr Schild coordinates.

Note $\psi =1$ is the event horizon of the Schwarzschild black hole. More generally the surface 
$\psi=1$ is easily verified to be a null submanifold of \eqref{ansatz0} for every choice of 
$\psi$ and $u$. This null manifold coincides with the event horizon of the \eqref{ansatz0} provided that 
$\psi$ and $u$ are chosen such that the metric \eqref{ansatz0} settles down into a collection of stationary 
black holes at late times.  Following 
\cite{Bhattacharyya:2015dva,Bhattacharyya:2015fdk} we refer to the submanifold $\psi=1$ as 
the membrane world volume. \footnote{Through this paper we assume that $\psi$ in \eqref{ansatz0} is chosen to 
ensure that the membrane surface is a smooth codimension one surface that is timelike when viewed 
as a submanifold of flat space (we have emphasized above that this surface is a null submanifold of 
the metric \eqref{ansatz0}). We also assume that 
$\psi$ is chosen to ensure that $\frac{1}{\psi^{D-3}}$ decays at spatial 
infinity.}

Note that as $\psi$ 
increases past unity $\frac{1}{\psi^{D-3}}$ decays to zero very rapidly. 
This decay is exponential in $D$ once $\psi-1 \gg \frac{1}{D}$. It follows 
that \eqref{ansatz0} represents a class of asymptotically flat spacetimes 
with the following property; the spacetime outside the event horizon deviates 
significantly from flat space only in a slab of thickness $\frac{1}{D}$ 
around the event horizon. We will refer to this as the membrane region. 

\cite{Bhattacharyya:2015dva,Bhattacharyya:2015fdk} set out to characterize 
solutions of the vacuum Einstein equations, $R_{MN}=0$, that reduce to 
metrics of the form \eqref{ansatz0} in the large $D$ limit, with corrections 
in a power series in  $\frac{1}{D}$. As we have reviewed above, when $\psi-1 \gg \frac{1}{D}$ the spacetimes \eqref{ansatz0} reduce to flat space.
Deviations from flatness are nonperturbatively small
in the $\frac{1}{D}$ expansion. Thus Einstein's equations are automatically
solved at all order in $1/D$ outside the membrane region. In order to 
obtain a true solution of Einstein's equations, the solution 
\eqref{ansatz0} needs to be corrected order by order in the $\frac{1}{D}$ 
expansion only in the membrane region.

Consider a region of size $\frac{1}{D}$ centered around any point $x_0$ on the 
event horizon of \eqref{ansatz0}. It may be shown that the metric of this ball is closely 
approximated by the metric in an equivalent small region centered around 
the appropriate event horizon point of {\it some} boosted Schwarzschild 
black hole provided that
\begin{equation}\label{ansatzt} 
\nabla^2 \left( \frac{1}{\psi^{D-3}} \right)=0, ~~~\nabla.u=0,
\end{equation}
(the contraction of all indices is achieved by use of the metric 
$\eta_{MN}$ in the equations above) \footnote{ When an expression like $\nabla^2$ acts  
on $\frac{1}{\psi^{D-3}}$ we get two distinct terms of order $D^2$ in two ways. 
The first term is $\propto (D-3)(D-2) \frac{(\nabla \psi)^2}{\psi^{D-1}}$. 
The second term is $\propto (D-3) \frac{\nabla^2 \psi}{\psi^{D-2}}$. Though 
the second term has one less explicit factor of $D$ than the first, 
it actually contributes at the same order in the $1/D$ expansion - i.e. 
at leading order - because of the contraction of indices in $\nabla^2$. 
This is the reason that \eqref{ansatz0} solves the leading order equations 
only if  $\nabla^2 \psi$ takes the same value as it does in a Schwarzschild 
black hole, leading to the first requirement listed in \eqref{ansatzt}. 
In a similar manner worldvolume
derivatives of the horizon shape and velocity field - which are of order 
unity - compete with derivatives acting on $\frac{1}{\psi^{D-3}}$ only if their 
order is enhanced by the contraction of a worldvolume index. 
The only first derivative expression involving the black hole velocity that 
has such a contraction is $\nabla.u$. It follows that \eqref{ansatz0} 
satisfies the leading order equations only if $\nabla.u$ takes the same value  
as it does on a Schwarzschild black hole. This leads to the second of 
\eqref{ansatzt}.}

. These equations need only be satisfied 
at leading order in $D$ and can be violated at subleading orders. 
As Schwarzschild black holes are exact solutions to 
Einstein's equations, it follows 
as a consequence that the spacetimes \eqref{ansatz0} {\it almost} 
solve Einstein's equations in the membrane region, provided that \eqref{ansatzt} 
is satisfied at every point on the membrane. 

The statement that Einstein's equations are `almost' solved in the membrane 
region has the following precise meaning. When evaluated in the membrane 
region the four derivative scalar $R_{AB} R^{AB}$ is in general of order $D^4$. 
This estimate follows immediately from the  fact that the metric 
varies on a length scale of order $1/D$ in the membrane region. Once we impose 
\eqref{ansatzt}, on the other hand, $R_{AB} R^{AB}$ turns out to be of order $D^2$, i.e.  In a coordinate system in which all components of the metric 
are of order unity, $R_{AB}$ is of order $D$; one order lower than the 
generic order suggested by a dimensional estimate.  
In other words \eqref{ansatzt} ensures that Einstein's equations are obeyed to 
leading order - but are generically violated at first subleading order. 
Consequently the metrics 
\eqref{ansatz0} - with the conditions \eqref{ansatzt} imposed at leading order- are plausible 
starting points for the construction of true 
solutions of Einstein's equations in a power series in $\frac{1}{D}$. 

The authors of \cite{Bhattacharyya:2015dva,Bhattacharyya:2015fdk} were able to 
carry out this perturbative expansion to first subleading order in $\frac{1}{D}$
(see below for a review). Interestingly they discovered that 
arbitrary metrics of the form \eqref{ansatz0} could {\it not} be corrected
to yield regular solutions to Einstein's equations at next order in 
$\frac{1}{D}$. It turns out to be possible to correct 
\eqref{ansatz0} at first order in $1/D$ only when the fields $\psi$ 
and $u$ obey an integrability constraint - a membrane equation of motion - 
 that we will describe in considerable detail below. Whenever this condition 
is obeyed, a regular correction (of order $1/D$) to the metric \eqref{ansatz0} 
was found in  \cite{Bhattacharyya:2015dva,Bhattacharyya:2015fdk}. 
The corrected metric obeys $R_{AB}={\cal O}(1)$ 
\footnote{More precisely, $R_{AB}={\cal O}(1)$ in coordinates in which 
all metric components are of order unity. More generally,  
$R_{AB} R^{AB}$ is of order unity. } ; 
i.e. once the corrections are taken into account, 
Einstein's equations are solved at leading {\it and first subleading order} 
in $\frac{1}{D}$. 

We now turn to a description of the integrability constraints mentioned 
in the previous paragraph.  Consider the surface 
$\psi=1$, viewed 
as a submanifold of flat space with metric $\eta_{MN}$; we refer to this 
submanifold as the membrane. Let $K_{MN}$ represent
the extrinsic curvature of this (generically timelike) submanifold.
Recall also that the velocity oneform field $u_M$ on the membrane surface 
is tangent to the membrane and so may be regarded as a oneform field in the 
membrane world volume. The authors of \cite{Bhattacharyya:2015dva,Bhattacharyya:2015fdk} found that the metric \eqref{ansatz0} 
could be corrected to a regular \footnote{By a regular solution we mean a 
solution with a smooth event horizon that is regular everywhere outside 
the event horizon.}  solution of Einsteins equations 
at first order if and only if the following constraints are obeyed 
\begin{equation} \label{VE1copy}
\left(\frac{\nabla^2 u_{A}}{\mathcal{K}}- \frac{\nabla_A\mathcal{K}}{\mathcal{K}}+ u_{C}K^C_A-u.\nabla u_A \right)\mathcal{P}^A_B=0
\end{equation}
where $\mathcal{P}^A_B=\delta^A_B + u^A u_B$ is the projector orthogonal 
to the velocity vector on the membrane world volume, and all covariant 
derivatives are taken with respect to the induced metric on the
membrane. The quantity $\mK$ is the trace of the extrinsic curvature of 
the membrane worldvolume.

The integrability conditions \eqref{VE1copy} have an interesting interpretation. 
They may be thought of as a set of $D-2$ equations for $D-2$ variables 
(one of these variables is the shape of the membrane, and the other $D-3$ 
variables are the components of the unit normalized, divergence free 
velocity field). In other words the equations \eqref{VE1copy} define an 
initial value problem for membrane dynamics. As every configuration that 
obeys \eqref{VE1copy} gives rise to a metric that obeys Einstein's equations to 
the appropriate order in $1/D$, it follows that solutions of the membrane 
equations \eqref{VE1copy} are in one to one correspondence with asymptotically 
flat dynamical black hole configurations that solve Einstein's equations 
to first subleading order in $1/D$. 

\subsection{The membrane paradigm at higher orders in $1/D$}

In this paper we demonstrate that first order perturbative procedure outlined above
extends systematically to arbitrary orders in the expansion in $\frac{1}{D}$. We will now 
very briefly outline our inductive argument. We assume that the perturbative procedure 
has been implemented upto $n^{th}$ order, i.e. that corrections to the metric 
\eqref{ansatz0} have been determined upto $n^{th}$ order in the $1/D$ expansion
in such a manner that $R_{MN}$ evaluated on the corrected solution is of order 
$D^{1-n}$. We then add further corrections of order $1/D^{n+1}$ to the metric 
(see \eqref{pertthy} and \eqref{metcorr}). At order $D^{n-1}$ we demonstrate that the 
Einstein constraint equations are independent of 
the new unknown correction functions when evaluated on the event horizon 
$\psi=1$. These equations determine the correction 
to the membrane equations (and the divergence condition on the velocity) at 
order $1/D^{n+1}$. Moving away from the horizon we argue that the order $D^{1-n}$ part of 
$R_{MN}$ takes the form listed in table \ref{Rbasisex}. Setting the expressions 
in this table yields a set of inhomogeneous linear differential equations that can be used 
to determine order $1/D^{n+1}$ corrections to the metric. Explicit expressions 
for the sources in these differential equations can only be obtained 
by grinding through the perturbative procedure, but we use 
a contracted Bianchi identity to demonstrate that the sources that occur in these equations 
are not all independent, but obey certain relations (see \eqref{bianchisrc}) 
at every order of perturbation theory. 
Using these relations we are able to integrate the inhomogeneous differential 
equations for any source functions and obtain an explicit and unique expressions 
for the metric corrections at order $1/D^{n+1}$ (see Section \ref{pfo}) that 
are manifestly regular and obey all required boundary conditions. 

As an illustration of the general method outlined above we explicitly 
implement the perturbative procedure to second subleading order in 
$\frac{1}{D}$. We find that the modified membrane equations take the form 
\begin{eqnarray}\label{VE2copy}
\nonumber&&\Bigg[\frac{\nabla^{2}u_{A}}{\mathcal{K}}-\frac{\nabla_{A} \mathcal{K}}{\mathcal{K}}+u^{B} K_{BA}-u\cdot\nabla u_{A}\Bigg]\CP^{A}_{C} \\\nonumber&+& \Bigg[\left(-\frac{u^{C}K_{CB}K^{B}_{A}}{\mathcal{K}}\right)+\left(\frac{\nabla^2\nabla^2 u_{A}}{\mathcal{K}^3}-\frac{u \cdot \nabla \mathcal{K}\nabla_{A}\mathcal{K}}{\mathcal{K}^3}-\frac{\nabla^{B}\mathcal{K}\nabla_{B}u_{A}}{\mathcal{K}^2}-2\frac{K^{CD}\nabla_{C}\nabla_{D}u_{A}}{\mathcal{K}^2}\right) \\ \nonumber &+&\left(-\frac{\nabla_{A}\nabla^2 \mathcal{K}}{\mathcal{K}^3} +\frac{\nabla_{A}\left(K_{BC}K^{BC}\mathcal{K}\right)}{\mathcal{K}^3}\right) + 3\frac{(u\cdot K\cdot u)(u\cdot \nabla u_{A})}{\mathcal{K}}-3\frac{(u\cdot K\cdot u)(u^{B} K_{BA})}{\mathcal{K}}\\  &-&6\frac{(u \cdot \nabla \mathcal{K})(u\cdot \nabla u_{A})}{\mathcal{K}^2}+6\frac{(u \cdot \nabla \mathcal{K})(u^{B}K_{BA})}{\mathcal{K}^2} + \frac{3}{(D-3)} u \cdot \nabla u_{A}-\frac{3}{(D-3)} u^{B}K_{BA} \Bigg]\CP^{A}_{C}=0\nonumber\\
\end{eqnarray}
while the divergence free condition on the velocity field is modified, at second subleading order, to the equation  
\begin{equation}\label{Divu}
 \nabla\cdot u=\frac{1}{2\mK}\left(\nabla_{(A}u_{B)}\nabla_{(C}u_{D)}\CP^{BC}\CP^{AD}\right)
\end{equation}

Note that the first line in \eqref{VE2copy} is simply a 
rewriting of \eqref{VE1copy}; the 2nd-4th lines of this equations represent 
corrections to \eqref{VE1copy}. There is a well defined sense (see below) in 
which each of these correction terms is of order $\frac{1}{D}$ relative 
to the leading order terms in the first line. It follows that the 
equations \eqref{VE2copy} represent small corrections to the leading order
equations \eqref{VE1copy}. The first order corrected membrane equation 
of motion  \eqref{VE2copy} and \eqref{Divu} are the main result of this paper. 

We then present explicit expressions for the second order sources
for all the inhomogeneous differential equations (see table \ref{Rbsrc2}). 
Plugging these sources into the general equations for the metric corrections 
at any order we obtain explicit results for the second order correction to 
the spacetime metric dual to any particular solution of the membrane equations 
of motion. 

The second order corrected membrane equations \eqref{VE2copy} admit a simple solution; a spherical 
membrane at rest. This solution is dual to the Schwarzschild black hole. 
As a check of our second order corrections to the membrane equations we use 
\eqref{VE2copy} to compute the spectrum of small fluctuations 
about this simple solutions. This spectrum is easy to obtain, and turns 
out to be in perfect agreement with the second order corrected spectrum 
of quasinormal modes obtained by Emparan Suzuki and Tanabe in 
\cite{Emparan:2014aba}, providing confidence in the correctness of 
\eqref{VE2copy}.

\section{Perturbation theory: general structure} \label{pertgen}

\subsection{A more detailed description of the starting ansatz} \label{md}

As we have explained in the introduction, the starting point of our perturbative construction 
of large $D$ solutions to Einstein's equations is the metric \eqref{ansatz0}. In the introduction 
we noted that the metrics \eqref{ansatz0} are parameterized by the $D$ dimensional function $\psi$ 
and the oneform field $u$. We assume these fields have a good large $D$ limit, 
i.e. that the length scale of variation in $\psi$ and $u$ is of order unity. 
Following 
\cite{Bhattacharyya:2015dva,Bhattacharyya:2015fdk}, however, consider  
two different functions $\psi$ with the same membrane surface (i.e. with coincident zero sets for $\psi -1$). These two functions define metrics 
\eqref{ansatz0} that coincide (outside the event horizon) at leading order 
in $1/D$ but differ at subleading orders in $1/D$. 
Similarly $u$ functions that agree on the membrane 
but differ off it lead to metrics \eqref{ansatz0} that differ only at subleading order in $1/D$. 

Any two metrics \eqref{ansatz0} that differ only at subleading orders in $1/D$ constitute equivalent 
starting points for the perturbative construction of solutions in the following sense: the end 
result of perturbation theory starting from the two different starting points will be the same. 
In order to construct all distinct final metrics we need only consider one member of each `equivalence
class' of metrics \eqref{ansatz0}. As explained above the equivalence classes are labeled by 
the zero set of the function $\psi -1$ (the membrane world volume) and the value of the velocity 
field on the membrane world volume. In order to pick a representative from each equivalence class that 
we can use to set up our perturbation theory we invent an arbitrary way of constructing the full 
function $\psi$ from its zero set, and the full velocity field $u$ from its values on the membrane. 
Following \cite{Bhattacharyya:2015dva,Bhattacharyya:2015fdk} we refer to the (essentially arbitrary) 
rule for achieving this construction as a subsidiary condition on the functions $\psi$ and $u$. 

For technical reasons, in this paper we utilize the subsidiary conditions of \cite{Bhattacharyya:2015dva} rather than that of \cite{Bhattacharyya:2015fdk}. We now describe these conditions in detail. 

Consider a given timelike membrane submanifold in flat space. At each point on the manifold consider a 
geodesic that shoots outwards from the manifold along its normal vector. The 
resultant collection of curves \footnote{These `curves' are actually straight 
lines as they are all geodesics in flat space. We use the term `curve' to bring
to mind the obvious generalization of this construction when the membrane is 
embedded in a curved spacetime.}
 is a 
spacefilling congruence of spacelike geodesics; caustics of this congruence, if any, only occur
at distances of order unity (rather than $1/D$) away from the membrane. 
\footnote{The quantity $\frac{D}{{\mathcal K}}$ gives a rough estimate for the 
distance away from the membrane at which the geodesics caustic. Below we 
explain that ${\mathcal K}$ is of order $D$ so that this caustic length 
scale is of order unity. }
We define the scalar function 
$B$ in the neighborhood of the membrane as follows; $B$ at any point is defined to be the signed proper distance, along the geodesic that passes through it, to the membrane. This distance is defined to be positive 
outside the membrane and negative inside the membrane. Note that $B$ vanishes on the membrane. We define 
\begin{equation}\label{ndef}
n_M=\nabla_M B
\end{equation}
It follows from our construction above that 
\begin{equation} \label{nnorm} 
n.n=1
\end{equation}
$n_A$ is the normal oneform to surfaces of constant 
$B$. We use the symbol $K_{MN}$ denote the extrinsic curvature of surfaces of constant $B$. Note 
of course that $n^AK_{AB}=0$. We also define ${\mathcal K}=K_A^A$. We then proceed to define the 
function $\psi$ as 
\begin{equation}\label{psidef}
\psi= 1 + \frac{{\mathcal K} B}{D-3} 
\end{equation}

In a similar manner we use the velocity function on the membrane to define a velocity oneform 
field in spacetime simply by parallel transport along our congruence of geodesics. It follows from 
our definitions above that 
\begin{equation}\label{auxeq} \begin{split}
&n.\nabla n_A=0 \\
& n.\nabla u_A=0
\end{split}
\end{equation}
The first line of \eqref{auxeq} follows upon differentiating 0
\eqref{nnorm}, using \eqref{ndef} and interchanging derivatives. 
This equation is in fact simply the geodesic equations 
for the congruence of geodesics that defines $B$. 
The equation on the second line of 
\eqref{auxeq} follows from the fact that $u$ is defined off the membrane 
 by parallel transport. It follows from \eqref{auxeq} that  
\begin{equation}\label{kab}
K_{AB}= (\eta_A^C-n_A n^C)\left(  \nabla_C n_D \right) 
\left( \eta^D_B -n^D n_B \right)  
= \left( \nabla_A - n_A (n. \nabla) \right) n_B 
= \nabla_A n_B= \nabla_A \nabla_B B
\end{equation}

Note that our definition of $n_A$ in this section, and the rest of this paper, 
differs slightly from the definition given 
in the introduction. The two definitions agree at leading order (which was all that was 
required in the discussion around \eqref{ansatz0} ) but differ at subleading orders in $1/D$.  The 
vector $n_A$ defined in this section - rather than the normal vector defined 
in the introduction - will be used through the rest of this paper. 

Using \eqref{psidef} it is easily verified that on the submanifold $B=0$
\begin{equation}\label{psicond} \begin{split}
&\psi \nabla^2 \psi= \frac{{\mathcal K}^2}{D-3} + 2 \frac{n.\nabla \mK}{D-3}\\
&(D-2) \nabla \psi. \nabla \psi =\frac{D-2}{D-3}\frac{{\mathcal K}^2}{D-3}
\end{split}
\end{equation} 
As we explain below, in the large $D$ limit taken in this paper $2 \frac{n.\nabla \mK}{D-3}$ 
is of order unity while $\frac{{\mathcal K}^2}{D-3}$ is order $D$. It follows that to leading order 
in $D$
$$ (D-2) \nabla \psi. \nabla \psi= \psi \nabla^2 \psi, ~~~i.e. 
\nabla^2 \left( \frac{1}{\psi^{D-3}} \right) =0$$
In other words our construction satisfies the first equation of \eqref{ansatzt}. We satisfy the second
equation in \eqref{ansatzt} by construction; we simply choose our $u$ oneform on the membrane such that its divergence vanishes at leading order in $D$. 
The divergence of $u$ will turn out not to vanish at a subleading order.

\subsection{Coordinate Choice for the correction metric}

In this paper we search for solutions of Einstein's equations in a power series expansion in 
$\frac{1}{D}$   
\begin{equation} \label{pertthy} 
\begin{split}
G_{MN}&= \eta_{MN} + h_{MN},\\
h_{MN}&= \sum_{n=0}^\infty \frac{h_{MN}^{(n)}}{(D-3)^n}\,,\\
\text{with, }h_{MN}^{(0)}&= \frac{O_M O_N}{\psi^{D-3}} ,\\
\end{split}
\end{equation}
Here 
\begin{equation}\label{Odef}
O_M=n_M-u_M
\end{equation}

We fix coordinate redefinition ambiguities by demanding 
\begin{equation}\label{corrcon}
h_{MN}O^N=0,
\end{equation}

Consider any point in the metric \eqref{ansatz0}. The tangent space built about 
this point has two special vectors; the vector $n$ and the vector $u$. All the 
other $D-2$ directions orthogonal to $n$ and $u$ are  equivalent and can be 
rotated into each other. It is thus useful to parameterize the most general 
fluctuation field $h_{MN}$ (subject to the gauge condition 
\eqref{corrcon})  in the form 
\begin{equation}\label{metcorr}
\begin{split}
h_{MN}^{(n)} &= H^{(S,n)} O_M O_N +O_{(M} H_{N)}^{(V,n)} +H^{(T,n)}_{MN} +   \frac{1}{D-3} H^{(Tr,n)} {\cal P}_{MN},\\
\text{where,}&\\
{\cal P}_{MN}= &\eta_{MN}-O_M n_N -O_N n_M + O_M O_N,\\
O^{N}  H_{N}^{(V,n)}&=0,~~~n^{N} H_{N}^{(V,n)}=0, ~~~O^{M} H^{(T,n)}_{MN} =0,~~~n^{M} H^{(T,n)}_{MN} =0, ~~~{\cal P}^{MN}H^{(T,n)}_{MN}=0,
\end{split}
\end{equation}
The superscripts $S$, $V$ and $T$ stand for scalar, vector and tensor 
respectively, and denote the transformation properties of the relevant 
symbol under the $SO(D-2)$ rotations in tangent space that leave $n$ and 
$u$ fixed. The superscript $Tr$ stands for trace, and labels a second scalar.

\subsection{Orders of $D$}

As we have explained above, in this paper we solve Einstein's equations 
in a systematic expansion in $\frac{1}{D}$. In order for this process to 
be well defined, we need to be able to unambiguously estimate the scaling 
with $D$ of various terms that appear in the metric and in the membrane 
equation of motion. Such an estimation is only unambiguous within 
subclasses of solutions, as we will now explain with an example. 

Consider a membrane whose world volume is a $D-2$ sphere (of radius $R$) 
times time. The trace of extrinsic curvature, ${\mathcal K}$, 
of this surface is easily shown to be $\frac{D-2}{R}$ and so is of order 
$D$ (assuming $R$ is of order unity). On the other hand 
the surface $S^p \times R^{D-2-p}$ times time has 
${\mathcal K}=\frac{p}{R}$. If $p$ and $R$ are both held fixed as $D$ is taken 
to infinity, ${\mathcal K}$ is of order unity for this surface. It follows 
that ${\mathcal K}$ cannot unambiguously be assigned a scaling with $D$
without making further assumptions. The same holds true of various other 
quantities (e.g. $\nabla^2 u_M$) that enter the metric and equation of motion. 

In this paper we follow \cite{Bhattacharyya:2015dva,Bhattacharyya:2015fdk}
and estimate the $D$ scalings of all terms as follows. We assume that 
\begin{itemize} 
\item Our starting ansatz is constructed by sewing together bits of the 
event horizon of black holes of radii $R$ and timelike velocity $u^M$ where 
$R$ and $u^M$ are everywhere finite and of order unity.
\item Our starting configuration (and so our full solution) preserves 
an $SO(D-p-2)$ rotational invariance with $p$ held fixed as $D$ is taken 
to infinity
\end{itemize}
As explained in \cite{Bhattacharyya:2015fdk}, these assumptions unambiguously 
specify the scaling with $D$ of all quantities of interest (in particular 
they force ${\mathcal K}$ to be of order $D$). 

We emphasize that in this paper we use the assumptions listed above only 
to estimate the scalings of $D$ of various quantities. When the assumptions 
listed in the previous paragraph are obeyed, the membrane equations 
and metrics listed in this paper certainly apply. However the formulae 
of this paper apply more generally to any spacetime whose variables 
scale with $D$ in the same manner in which they would if the assumptions 
above were obeyed - a much larger class of configurations. 

\subsection{All orders definition of the membrane surface and velocity} 
\label{sv}

As explained in subsection \ref{md}, the metric \eqref{ansatz0} - the starting
point of our perturbative expansion - is completely determined by the 
shape of a membrane and a velocity field on the membrane. To what precision 
can this procedure be reversed? In other words if we are given a solution 
to Einstein's equations of the appropriate kind, how precisely can we 
read off the corresponding `shape' and `velocity' of the membrane?

We could attempt to identify the membrane shape and velocity field by 
simply expanding the exact solution in powers of 
$1/D$ and focusing attention on the leading order term. By comparing with \eqref{ansatz0} we could then read off 
the membrane shape and velocity field. While this procedure is simple, 
a moment's thought will convince the reader that it is ambiguous at 
all orders in $1/D$ save the leading order. \footnote{For instance, the 
velocity redefinition $u^\mu \rightarrow u^\mu + \delta u^\mu/D$ does not 
change the metric at leading order in $1/D$.} In other words the requirement 
that our solution reduce to \eqref{ansatz0} defines the membrane shape and 
velocity only at leading order, leaving the subleading corrections to 
these quantities ambiguous. In this subsection we will fix this ambiguity  
by adopting a more precise definition of the shape and velocity field. This 
definition agrees with that of \eqref{ansatz0} at leading order, but 
is precise at all orders. We use this precise definition 
in the computations presented in the rest of this paper. 

We define the membrane shape to be the location of the event horizon of our 
spacetime, and will choose higher order corrections to the metric 
\eqref{ansatz0} to ensure that this event horizon coincides with the 
surface $\psi=1$. 

Turning to the velocity field, let $G^{AB}$ denote the full spacetime
inverse metric. Let $n_A$ be the oneform normal to the event horizon. 
We define the velocity field on the membrane by the requirement that 
\begin{equation}\label{vfm}
u^A= G^{AB}n_B
\end{equation}
on the event horizon (i.e. at $\psi=1$). In other words the velocity field 
is a tangent vector to the generators of 
the event horizon. It is easily verified that \eqref{vfm} is 
a true equation for the starting point of perturbation theory 
\eqref{ansatz0}. We will choose corrections to the perturbative ansatz 
to ensure that \eqref{vfm} holds at all orders in $1/D$. 

The requirement \eqref{vfm} together with the requirement that $\psi=1$ is the 
exact event horizon of our spacetime are easily seen to be 
satisfied provided that 
\begin{equation}\label{velhor}
\begin{split}
H^{(S)}(\psi=1)&=0\\
H^{(V)}_M(\psi=1)&=0\\
\end{split}
\end{equation}
The first condition ensures that $G^{MN}\partial_M \psi \partial_N \psi =0$,
i.e. $d\psi$  is null at $\psi=1$ while the second 
condition then ensures that the full spacetime metric on the event horizon 
takes the form 
$$ \eta_{MN} + O_M O_N+  H^{(T)}_{MN} + \frac{1}{D-3} H^{Tr} {\CP}_{MN}$$
Let us write this metric in a the local basis of oneforms 
$(n, u, Y_{a})$ where $Y_a$ is any $D-2$ dimensional basis of oneforms
chosen orthogonal to $n$ and $u$. In this basis the metric takes a block 
diagonal form with a $2 \times 2$ block (with basis $n$ and $u$) and a 
$D-2 \times D-2$ block (with basis $Y_a$). It follows that the inverse metric 
also has this block diagonal structure. Note that the $2 \times 2$ 
block is universal, i.e. it is the same at every order in perturbation theory. 
This block is the only one that contributes in \eqref{vfm}. As 
\eqref{vfm} holds at leading order, it follows 
that the conditions \eqref{velhor} ensure that \eqref{vfm} holds at every 
order in perturbation theory.

Recall that according to \eqref{ansatzt} the velocity field used in 
\eqref{ansatz0} is divergence free at leading order in $\frac{1}{D}$. 
As we will see below, the divergence of the velocity field defined in this 
subsection will not, in general, vanish at subleading orders in $1/D$. 

\subsection{Structure of the equations of perturbation theory}

Our perturbative procedure proceeds as follows. We assume that our 
solution takes the form \eqref{pertthy} together with 
\eqref{corrcon} and \eqref{metcorr}. The Ricci tensor of this metric - 
evaluated in a slab of spacetime of thickness $1/D$ around $\psi=1$ - 
takes the schematic form 
\begin{equation}\label{rt}
R_{MN}= \sum_n D^{2-n} R_{MN}^{n} 
\end{equation}

Let us imagine that we have implemented our perturbative procedure to order 
$n-1$, i.e. that we have determined $h^{(m)}_{MN}$ for $m=1 \ldots n-1$ in a 
manner that ensures that $R_{MN}^{(m)}=0$ for $m= 0 \ldots n-1$.   
In order to go to one higher order in perturbation theory we must solve 
for $h_{MN}^{(n)}$ to ensure that $R_{MN}^n$ also vanishes.  

Schematically  
$$ R_{MN}^{(n)}= C_{MN}^{PQ} h^{(n)}_{PQ} + {\mathcal S}^{(n)}_{MN}$$
where  $C_{MN}^{PQ}$ is a linear differential operator with derivatives only in 
the $\psi$ direction and ${\mathcal S}^{(n)}_{MN}$ is a source function. 
As  $h^{(n)}_{PQ}$ is already of order $n$, the differential operator
$C_{MN}^{PQ}$ is built entirely out of the zero order background metric 
\eqref{ansatz0}, and so is the same at every order. On the other 
hand the source function ${\mathcal S}^{(n)}_{MN}$ is proportional to expressions
of $n^{th}$ order in $1/D$ built out of derivatives of the membrane velocity 
and shape function, and is different at every order. 

At every point of the event horizon of the ansatz metric \eqref{ansatz0}
there are two distinguished vectors; $n^A$ and $u^A$. Let 
$${\cal P}_{AB}= \eta_{AB} -n_A n_B + u_A u _B$$
denote the projector orthogonal to these two vectors (all dot products 
taken in flat space). Instead of dealing directly with the components of 
$R_{MN}$ we find it more convenient to use a basis adopted to $u^A$ and $n^A$ listed in table \ref{Rbasis}.
 
\begin{center}
	\begin{table}[h]
		\caption{Basis of components of $R_{MN}$}\label{Rbasis}
		\begin{tabular}{ |c|c|c| }
			\hline
			Scalar sector  & Vector sector & Tensor sector \\ 
			\hline
			$R^{S_1} = O^M R_{MN} O^N $& $ R^{V_1}_{L}= O^{M}R_{MN}\CP^{N}_{L} $ & $R^{T}_{AB} = \CP^M_A R_{MN} P^N_B - \frac{\CP_{AB}}{D-2}\CP^{MN}R_{MN} $ \\ 
			$R^{S_2} = O^M R_{MN} u^N $& $R^{V_2}_{L}= u^{M}R_{MN}\CP^{N}_{L}$& $~$ \\
			$ R^{S_3} = u^M R_{MN} u^N $ & $~$     & $~$     \\
			$R^{S_4}= R_{MN}\CP^{MN} $& $~$ &  \\
			\hline
		\end{tabular}
	\end{table}
\end{center}
\noindent

By explicit computation (plugging \eqref{pertthy} into the formula for  
the Ricci tensor) we find that the linear combinations listed in 
Table \ref{Rbasis}  of the curvature components $R^n_{MN}$ (see \eqref{rt}) 
are given by the expressions listed in Table \ref{Rbasisex}. 

 In table \ref{Rbasisex}, fluctuation fields $H^S$, $H^{Tr}$ $H^V_A$ and $H^T_{MN}$ are taken 
to be of $n^{th}$ order and all source functions 
(e.g. ${\mathcal S}^{S_1}$) also understood to be $n^{th}$ order sources.
All appearances of $\nabla.u$  
\footnote{$\nabla.u$ is the divergence of the velocity field thought of 
as a vector field in $R^{D-1,1}$. On the surface $\psi=1$, however, 
$\nabla.u$ coincides with the membrane worldvolume divergence of 
velocity field  (this follows upon using the second of \eqref{auxeq}).}
in the table \ref{Rbasisex} should also be understood 
as follows. Naively $\nabla.u$ is of order $D$. For that reason we expand  
\begin{equation}\label{nu}
\nabla.u=(D-3) \left( \sum_{n=0}^\infty \frac{(\nabla.u)_n}{(D-3)^{n}} \right)
\end{equation}
Every appearance of $\nabla.u$ in table \ref{Rbasisex} should actually 
be replaced by $(\nabla.u)_n$. We have already seen in the introduction 
that $(\nabla.u)_0=0$. We will see below that $(\nabla.u)_1$ also vanishes, 
but that $(\nabla.u)_2$ is nonzero. 

\begin{center}
	\begin{table}[t]
		\caption{Expressions for basis of $R_{MN}$}\label{Rbasisex}
		\resizebox{\columnwidth}{!}{
		\begin{tabular}{ |c| }
			\hline
			Scalar sector  \\ 
			\hline
			$R^{S_1} = \left( \frac{-\mK^2}{2(D-3)^2} \right) \frac{d^2 H^{(Tr)}}{dR^2}+{\mS}^{S_1}(R)  $\\ 
			$R^{S_2} = \left( \frac{\mK^2}{2(D-3)^2} \right) e^{-R}\frac{d}{dR}\left(e^R\frac{d}{dR}H^{(S)}\right) - \frac{\mK^2}{4(D-3)^2}e^{-R}\frac{d}{dR}H^{(Tr)} + \frac{\mK}{2(D-3)}\nabla^MH^{(V)}_M $\\$+ {\mS}^{S_2}(R) + \frac{\mK}{2(D-3)}e^{-R}~\nabla.u $\\
			$R^{S_3} = \left( \frac{\mK^2}{2(D-3)^2} \right)e^{-2R}(1-e^R)\frac{d}{dR}(e^R\frac{dH^{(S)}}{dR}) $\\$ - \left( \frac{\mK^2}{4(D-3)^2} \right)e^{-2R}(1-e^R)\frac{dH^{(Tr)}}{dR} - \frac{\mK}{2(D-3)}e^{-R}\nabla^MH^{(V)}_M + {\mS}^{S_3}(R) + \frac{\mK}{2(D-3)}~e^{-2R}~\nabla.u$ \\
			$R^{S_4} = \left( \frac{\mK^2}{(D-3)^2}\right) e^{-R}\frac{d}{dR}(e^R H^{(S)}) + \left( \frac{\mK^2}{2(D-3)^2} \right)e^{-2R}(1-e^R)\frac{d}{dR} \left( e^R \frac{d}{dR}H^{(Tr)} \right) $\\$ - \left( \frac{\mK^2}{2(D-3)^2} \right) \frac{dH^{(Tr)}}{dR}  + \frac{\mK}{D-3}\nabla^M H^{(V)}_M + \frac{2\mK}{D-3}\frac{d}{dR}\nabla^M H^{(V)}_M + \nabla^M \nabla^N H^{(T)}_{MN} + {\mS}^{S_4}(R)-~\frac{\mK}{(D-3)} e^{-R}\nabla.u$ \\
			\hline
			Vector sector  \\ 
			\hline
			$R_M^{V_1} =\left( \frac{\mK^2}{2(D-3)^2} \right) e^{-R} \frac{d}{dR}(e^{R} \frac{d}{dR} H^{(V)}_{M}) + \frac{1}{2}\frac{\mK}{(D-3)} \frac{d}{dR} \left(  \nabla^N H^{(T)}_{NM} \right)+{\mS}^{V_1}_{M}(R) $ \\ 
			$R_M^{V_2} = \left( \frac{\mK^2}{2(D-3)^2} \right) e^{-2R}(1- e^{R}) \frac{d}{dR}(e^{R} \frac{d}{dR} H^{(V)}_{M}) + {\mS}^{V_2}_{M}(R)$\\
			\hline
			Tensor sector  \\ 
			\hline
			$R^{T}_{AB} = \left( \frac{-\mK^2}{2(D-3)^2} \right) e^{-R} \frac{d}{dR}\left(\left(e^R-1\right) \frac{dH^{(T)}_{AB}}{dR}\right)   + {\mS}^{T}_{AB}(R) $ \\ 
			\hline
		\end{tabular}}
	\end{table}
\end{center}
\noindent

In order to obtain Table \ref{Rbasisex} we have worked in the neighbourhood 
of the surface 
$\psi=1$ and the variable $R$ is defined by $R=(D-3)(\psi-1)$.
\footnote{We will explain below that the sources listed in  Table \ref{Rbasisex}
are not completely independent, but are constrained by the well known relation 
\begin{equation}\label{bi}
\nabla^M \left( R_{MN} - \frac{\tilde{R}}{2}G_{MN} \right) = 0 
\end{equation} .}

\subsection{The Einstein Constraint Equations}

In the process of solving for the fluctuation fields $h_{MN}^{(n)}$ we 
will find the Einstein constraint equations (relevant to the 
foliation of our spacetime in slices of constant $\psi$) particularly useful. 
We will now provide a careful definition of these equations. 

Let us define 
\begin{equation} \label{eeqs}
E_{MN}\equiv R_{MN}-\tilde{R} \frac{G_{MN}}{2}
\end{equation}
where $\tilde{R}$ is the Ricci scalar.
The constraint equations are defined by the relations
\begin{equation}\label{EEconstr}
 E^{(ec)}_M = E_{MN}G^{NL}n_L        
\end{equation}
We have a total of $D$ constraint equations. These equations decompose into two scalars and one 
vector under local $SO(D-2)$ rotations. 

Let us imagine we have solved for our membrane metric at $(n-1)^{th}$ order in perturbation
theory, and are now attempting to solve for the metric correction at $n^{th}$ order. 
If, in this process, we evaluate the constraint equation \eqref{EEconstr} and retain 
terms only up to $n^{th}$ order then we need use $G^{NL}$ on the RHS of \eqref{EEconstr}
only at zero order (i.e. from the metric \eqref{ansatz0}), because $E_{MN}$ is already of 
$n^{th}$ order. It follows that the $n^{th}$ order scalar and vector constraint equations 
are simply linear combinations of the $n^{th}$ order scalars and vectors listed in 
table \ref{Rbasis}. We will now determine the relevant linear combinations. In order to 
to this we first determine the $n^{th}$ order Ricci scalar ${\tilde R}$ as a linear combination 
of the scalars in table \ref{Rbasis}.
\begin{equation}\label{rs}
\tilde{R} = R_{AB}G^{AB}\\ =\left( R^{AB}P_{AB} + O.R.O(1 - e^{-R}) + 2O.R.u \right) 
 = (R^{S_4}+(1-e^{-R})R^{S_1}+2R^{S_2}) 
 \end{equation} 
Using this equation we find 
\begin{equation}\label{EEconflat}
\begin{split}
 E^{(ec)}_M &= \left( R_{MN}- \frac{\tilde{R}}{2}G_{MN} \right) G^{NL}n_L \\
 &= R_{MN}O^N(1-e^{-R}) + R_{MN}u^N -\frac{1}{2}\tilde{R}~ n_M  \\
\end{split}                                                                                                \end{equation}
By dotting \eqref{EEconflat} with $n$ and $u$ or by projecting it orthogonal to these vectors 
we finally obtain the $n^{th}$ order constraint equations written as linear combinations 
of the scalars and vectors in table \ref{Rbasis}.
 \begin{equation}\label{DCrel}
 \begin{split}
  &E^{S_1}=E^{(ec)}_M u^M=(1-e^{-R})R^{S_2}+R^{S_3} \\
  &E^{S_2}=E^{(ec)}_M O^M=\frac{1}{2}\left((1-e^{-R})R^{S_1}-R^{S_4}\right) \\
  &E^{V}_L=E^{(ec)}_N\CP^N_L=(1-e^{-R})R^{V1}_L+R^{V_2}_L
 \end{split}
 \end{equation}

 The explicit form of the $n^{th}$ order constraint equations is listed in table \ref{Ebasis}
below
 \begin{center}
	\begin{table}[h]
		\caption{Listing of constraint equations}\label{Ebasis}
		\resizebox{\columnwidth}{!}{
		\begin{tabular}{ |c| }
			\hline
			Vector constraint \\ 
			\hline
			$E^{V}_M = E^{(ec)}_N\CP^N_M =(1-e^{-R})R^{V_1}_M+R^{V_2}_M $\\$ = \frac{1}{2}\frac{\mK}{(D-3)}(1-e^{-R}) \frac{d}{dR} \left(  \nabla^A H^{(T)}_{AM} \right) + {\mathcal V}_M^{V}(R) $\\
			\hline
			Scalar constraint 1 \\ 
			\hline
			$E^{S_1} = E^{(ec)}_M u^M =(1-e^{-R})R^{S_2}+R^{S_3} $\\$ =\frac{\mK}{2(D-3)}(1-e^R)  \frac{d}{dR}\left( \nabla^M H^{(V)}_M \right) - \frac{\mK}{2(D-3)}e^{-R} \nabla^M H^{(V)}_M + {\mathcal V}^{S1}(R) + \frac{\mK}{2(D-3)}~e^{-R}~\nabla.u $\\
			\hline
			Scalar constraint 2 \\ 
			\hline
			$E^{S_2} = E^{(ec)}_M O^M =\frac{1}{2}\left((1-e^{-R})R^{S_1}-R^{S_4}\right)= -\frac{\mK}{2(D-3)} \frac{d}{dR}\left( \nabla^M H^{(V)}_M \right)-\frac{\mK}{(D-3)} \nabla^M H^{(V)}_M $ \\ $+ \frac{\mK^2}{4(D-3)^2}(2-e^{-R})\frac{d}{dR}H^{(Tr)}  - \frac{\mK^2}{2(D-3)^2}\left( \frac{d}{dR}H^{(S)}+H^{(S)} \right) - \frac{1}{2} \nabla_M\nabla_N H^{(T)}_{MN} + {\mathcal V}^{S2}(R) + \frac{\mK}{2(D-3)} e^{-R}~\nabla.u $\\
			\hline
		\end{tabular}}
	\end{table}
\end{center}
\noindent 
As in table \ref{Rbasis}, all fluctuation fields in table \ref{Ebasis} should be taken to be 
of $n^{th}$ order. The source functions in table \ref{Ebasis} are also of $n^{th}$ order 
and are given in terms of the sources in table \ref{Rbasis} and the as yet unknown quantity 
$\nabla.u$  by 
\begin{equation}\label{constsc}
 \begin{split}
  &{\mathcal V}^{S_1}(R) =(1-e^{-R}){\mathcal S}^{S_2}(R)+{\mathcal S}^{S_3}(R) \\
  &{\mathcal V}^{S_2}(R) =\frac{1}{2}\left[(1-e^{-R}){\mathcal S}^{S_1}(R)- {\mathcal S}^{S_4}(R)\right] \\
  &{\mathcal V}^{V}_L(R)=(1-e^{-R}){\mathcal S}^{V_1}_L(R)+{\mathcal S}^{V_2}_L(R)
 \end{split}
\end{equation}

Now it is well known that the Einstein tensor obeys the identity 
\begin{equation}\label{bianchi}
 \nabla_M E^{MN}=0
\end{equation}
It is also well known (and easy to see) that this identity ensures that the `normal' derivative 
of the constraint equations is a linear combination of the `in plane' derivatives of Einstein's 
equations. \footnote{This is the fact that ensures that if all Einstein constraint equations are solved on one 
`time' slice then they are automatically solved on the next `time' slice. In other 
words, in order to solve Einstein's equations you need only solve the constraint equations on one time 
slice provided you solve the other equations - lets call them the dynamical equations - everywhere.}
Within the perturbation theory of interest to this paper the equation \eqref{bianchi} may be evaluated and projected onto its scalar and vector sectors and shown to be equivalent to the following relations
 
\begin{equation}\label{bianchirel}
\begin{split}
 &\frac{d}{dR}E^{V}_M + E^{V}_M + \frac{(D-3)}{\mK} \nabla^N R^T_{NM}=0 \\ 
 &\frac{d}{dR}E^{S_1} + E^{S_1} + \frac{(D-3)}{\mK} \nabla^N R^{V_2}_{N}=0 \\ &\frac{d}{dR}E^{S_2} + E^{S_2} + \left( \frac{1}{2}R^{S_1} + R^{S_2} + \frac{1}{2}R^{S_4} \right) + \frac{(D-3)}{\mK} \nabla^N R^{V_1}_{N}=0
\end{split}
\end{equation}
Using \eqref{DCrel} the RHS of these relations may be recast in the equivalent form
\begin{equation}\label{biarel}
\begin{split}
 &\frac{d}{dR}E^{V}_M + (1-e^{-R})R_M^{V_1} + R_M^{V_2} + \frac{(D-3)}{\mK} \nabla^N R^T_{NM}=0 \\ &\frac{d}{dR}E^{S_1} + (1-e^{-R})R^{S_2} + R^{S_3} + \frac{(D-3)}{\mK} \nabla^N R^{V_2}_{N}=0 \\ &\frac{d}{dR}E^{S_2} + \frac{1}{2}e^{-R}R^{S_1} + (1-e^{-R})R^{S_1} + R^{S_2} + \frac{(D-3)}{\mK} \nabla^N R^{V_1}_{N}=0
\end{split}
\end{equation}
In either form these  equations express the $R$ derivatives of the Einstein constraint equations 
\eqref{DCrel} in terms of linear combinations of the Einstein equations. 
Using the explicit expressions in tables \ref{Rbasisex} and \ref{Ebasis}, it is possible to verify 
that the equations \eqref{bianchirel} are indeed obeyed, provided that the scalar and vector sources in
table \ref{Rbasisex} and \ref{Ebasis} are  not all independent but are constrained by the following relations
\begin{equation}\label{bianchisrc}
\begin{split}
 &\frac{d}{dR}{\mathcal V}^{V}_M + {\mathcal V}^{V}_M + \frac{(D-3)}{\mK} \nabla^N {\mathcal S}^T_{NM}=0 \\ 
 &\frac{d}{dR}{\mathcal V}^{S_1}+  {\mathcal V}^{S_1}  + \frac{(D-3)}{\mK} \nabla^N {\mathcal S}^{V_2}_{N}=0 \\ &\frac{d}{dR}{\mathcal V}^{S_2} + {\mathcal V}^{S_2} + \left[ \frac{1}{2}{\mathcal S}^{S_1} + \left( {\mathcal S}^{S_2}  +\frac{\mK}{2(D-3)}e^{-R}~\nabla.u \right) + \frac{1}{2}\left({\mathcal S}^{S_4}-\frac{\mK}{(D-3)}e^{-R}\nabla.u\right) \right] \\&+ \frac{(D-3)}{\mK} \nabla^N {\mathcal S}^{V_1}_{N}=0
\end{split}
\end{equation}

Note that we have two relations between the four scalar sources and one relation between the 
two vector sources in  table \ref{Rbasisex}. Note  that the relations also involve the as yet unknown 
quantity $\nabla.u$. Later in this paper we will explicitly verify that the sources that appear in the first and second 
order calculation obey the relations \eqref{bianchisrc}. However we would like to emphasize here 
that these relations are necessarily obeyed at every order in perturbation theory.

\subsection{Choice of basis for the constraint and dynamical equations}

Because we have the linear relationship between constraint and dynamical equations we use the following basis for solving the scalar, vector and tensor fluctuations

\begin{equation}\label{basisRE}
 \begin{split}
  &\text{Tensor:}~~~~ R^{T}_{AB} \\
  &\text{Vector:}~~~~ R^{V_2}_M,~~E^V_M \\
  &\text{Scalar:}~~~~ R^{S_1},~~R^{S_2},~~E^{S_1},~~E^{S_2} 
 \end{split}
\end{equation}

From now on we write every expression in this basis. The expressions that we get from Bianchi identities i.e. equations \eqref{bianchirel},\eqref{biarel} can be converted to the basis \eqref{basisRE} as

\begin{equation}\label{BiBa}
 \begin{split}
  &\frac{d}{dR}E^{V}_M + E^{V}_M + \frac{(D-3)}{\mK} \nabla^N R^T_{NM}=0 \\ 
 &\frac{d}{dR}E^{S_1} + E^{S_1} + \frac{(D-3)}{\mK} \nabla^N R^{V_2}_{N}=0 \\ 
 &\frac{d}{dR}E^{S_2} + (1-\frac{1}{2}e^{-R})R^{S_1} + R^{S_2} + \frac{1}{1-e^{-R}} \frac{(D-3)}{\mK}\nabla^M \left(  E_M^V- R^{V_2}_{M} \right) =0
 \end{split}
\end{equation}

The corresponding relationship between the sources is given by

\begin{equation}\label{BiBaSrc}
 \begin{split}
 &\frac{d}{dR}{\mathcal V}^{V}_M + {\mathcal V}^{V}_M + \frac{(D-3)}{\mK} \nabla^N {\mathcal S}^T_{NM}=0 \\ 
 &\frac{d}{dR}{\mathcal V}^{S_1}+{\mathcal V}^{S_1}   + \frac{(D-3)}{\mK} \nabla^N {\mathcal S}^{V_2}_{N}=0 \\  &\frac{d}{dR}{\mathcal V}^{S_2}  + (1-\frac{1}{2}e^{-R}){\mathcal S}^{S_1} + {\mathcal S}^{S_2}+ \frac{1}{1-e^{-R}}\frac{(D-3)}{\mK} \nabla^N \left({\mathcal V}^{V}_{N}-{\mathcal S}^{V_2}_{N}\right)=0
 \end{split}
\end{equation}

\section{Perturbation theory at first order} \label{pfo}

In this section we will explicitly solve for the first order correction metric $h^{(1)}_{MN}$. 
However we will perform our analysis in a manner that makes the generalization to higher 
orders obvious. 

\subsection{Listing first order source functions}
As we have explained in the previous section, the components of $R^1_{MN}$ are given in terms of 
$h^{(1)}_{MN}$ by the expressions in Table \ref{Rbasisex} with particular values for the source 
functions in that table. By explicit calculation at first order we find that these source functions
are given by the values listed in the table \ref{Rbsrc}. 
\begin{center}
	\begin{table}[h!]
		\caption{Sources of $R_{MN}$ equations at 1st order}\label{Rbsrc}
		\resizebox{\columnwidth}{!}{
		\begin{tabular}{ |c| }
			\hline
			Scalar sector  \\ 
			\hline
			${\mS}^{S_1}(R) = 0  $\\ 
			${\mS}^{S_2}(R) = \frac{\mK}{2(D-3)}e^{-R}u.K.u-\frac{e^{-R}(-1+R)}{2}\frac{u.\nabla\mK}{(D-3)} - \frac{\mK^2}{2(D-3)^2}e^{-R}(-3+2R)$\\
			${\mS}^{S_3}(R) = \frac{1}{2\mK(D-3)}Re^{-R}\nabla^2\mK-\frac{e^{-2R}(-2+2e^R+R)}{2}\frac{u.\nabla \mK}{(D-3)} + \frac{\mK^2}{2(D-3)^2}e^{-2 R} \left(3 e^R (R-1)-2 R+3\right) $ \\
			${\mS}^{S_4}(R) = e^{-R}(-1+R)\frac{u.\nabla\mK}{(D-3)} + \frac{\mK^2}{(D-3)^2}e^{-R}(-1+2R) $ \\
			\hline
			Vector sector  \\ 
			\hline
			 $ {\mS}^{V_1}_{A}(R) = \frac{\mK}{2(D-3)} e^{-R} \left( u^M K_{MN} - u^{M}\nabla_{M} u_{N} \right) \CP^N_A   $ \\
			$  {\mS}^{V_2}_{A}(R) = \frac{\mK}{2(D-3)} e^{-2R} \left( u^M K_{MN} - u^{M}\nabla_{M} u_{N} \right) \CP^N_A + \frac{e^{-R}}{2}\left( \frac{\nabla^2 u_A}{(D-3)}-\frac{\nabla_A \mK}{(D-3)}  \right) $\\
			\hline
			Tensor sector  \\ 
			\hline
			$ {\mS}^{T}_{AB}(R) = 0 $ \\ 
			\hline
		\end{tabular}}
	\end{table}
\end{center}
\noindent

Moreover the constraint equations take the form listed in Table \ref{Ebasis} with  first order source functions listed in Table \ref{Esrc}. 
We list the corresponding sources to the constraint equations at 1st order in table \ref{Esrc}.  We have verified that our explicit expressions for the sources obey the constraints \eqref{bianchisrc}.

We now proceed to solve the metric corrections at 1st order i.e. $h^{(1)}_{MN}$. We impose the conditions \eqref{velhor} as discussed in section \ref{sv}. 

 \begin{center}
	\begin{table}[t]
		\caption{Sources to constraint equations at 1st order}\label{Esrc}
		\begin{tabular}{ |c| }
			\hline
			Vector constraint source\\ 
			\hline
			$ {\mathcal V}_M^{V}(R) = \frac{e^{-R}}{2}\left( \frac{\nabla^2 u_M}{(D-3)}-\frac{\nabla_M \mK}{(D-3)} + \frac{\mK}{(D-3)}(u^AK_{AM}-u.\nabla u_M) \right) $\\
			\hline
			Scalar constraint 1 source\\ 
			\hline
			$ {\mathcal V}^{S_1}(R) = \frac{1}{2\mK(D-3)}Re^{-R}\nabla^2\mK - \frac{-e^{-2R}+e^{-R}(1+R)}{2}\frac{u.\nabla\mK}{(D-3)} $\\$+ \frac{\mK}{2(D-3)}e^{-R}(1-e^{-R})u.K.u + Re^{-R}\frac{\mK^2}{2(D-3)^2} $\\
			\hline
			Scalar constraint 2 source\\ 
			\hline
			$ {\mathcal V}^{S_2}(R) = \frac{e^{-R}}{2} \left( \frac{\mK^2}{(D-3)^2}(1-2R) + \frac{u.\nabla \mK}{(D-3)} (1-R) \right) $\\
			\hline
		\end{tabular}
	\end{table}
\end{center}
\noindent

\subsection{Tensor sector}

In this sector we have a single equation  for the single variable $H^{(T)}_{MN}$.
This equation is obtained by equating the last line of Table 
\ref{Rbasisex} to zero and takes the form 
\begin{equation} \label{teneq}
	\begin{split}
		R^{T}_{AB} = e^{-R} \frac{d}{dR}\left(\left(e^R-1\right) \frac{dH^{(T)}_{AB}}{dR}\right) \left( \frac{-\mK^2}{2(D-3)^2} \right)  + {\mS}^{T}_{AB}(R)=0
	\end{split}
\end{equation}
 where ${\mS}^{T}_{AB}(R)$ is the source for the tensor sector. At first 
order it turns out that ${\mS}^{T}_{AB}(R)=0$ (see Table \ref{Esrc}). 
In order to facilitate generalizations to higher orders however, 
in this subsection we will solve \eqref{teneq} for an arbitrary source 
function, and substitute ${\mS}^{T}_{AB}(R)=0$ only at the end of the
calculation. 

Integrating \eqref{teneq} once we find 
\begin{equation}\label{fi}
\begin{split}
\frac{d}{dR}( H^{(T)}_{AB})= \left( \frac{-2(D-3)^2}{\mK^2} \right) \frac{-1}{e^R-1}\int_0^R e^{x}{\mS}^{T}_{AB}(x) dx
\end{split}
\end{equation}
The condition that $H^{(T)}_{AB}$ (and so RHS 
of \eqref{fi}) is regular at $R=0$ fixes
the lower limit of the integral in  \eqref{fi}. 
Integrating a second time we find 
\begin{equation}\label{si}
\begin{split}
 H^{(T)}_{AB} &=\left( \frac{-2(D-3)^2}{\mK^2} \right) \int_R^{\infty}\frac{dy}{e^y-1}\int_0^y e^{x}{\mS}^{T}_{AB}(x) dx \\
 &= \left( \frac{2(D-3)^2}{\mK^2} \right) \bigg[ \log(1-e^{-R})\int_0^R e^x S^{T}_{AB}(x)dx + \int_R^{\infty} \log(1-e^{-x})e^x S^{T}_{AB}(x) \bigg]
\end{split}
\end{equation}
where the upper limit in the outer integral in \eqref{si} is fixed by the requirement that 
$H^{(T)}_{AB}$ decay at large $R$. 

In summary, the tensor fluctuation $H^{(T)}_{AB}$ is given at any order, in 
terms of the tensor source function ${\mS}^{T}_{AB}(x)$ at that order, by the 
expression \eqref{si}. Note that $H^{(T)}_{AB}$ is uniquely determined  
by its source function; requirements of regularity at $R=0$
and decay at infinity unambiguously fix all integration constants in 
\eqref{teneq}.  

As we have mentioned above, at first order ${\mS}^{T,1}_{AB}(R)=0$. 
It follows from \eqref{si} that the first order tensor fluctuation $H^{(T)}_{AB}$ also vanishes and so 
 \begin{equation} \label{foten}
  H^{(T,1)}_{AB}=0
 \end{equation}

\subsection{Vector Sector}

\subsubsection{Constraint Equation and the Membrane Equation of Motion}

In the vector sector we have two equations for the single 
variable $H^{(V)}_M$. The two equations may be chosen to be the vector 
constraint equation $E^V_M$ (see the first line of Table \ref{Ebasis}) and 
the equation $R_L^{V_2}=0$ (see Table \ref{Rbasisex}).

One cannot, of course, solve two equations for a single variable unless
one linear combination of the two equations is an identity. Indeed the first equation of 
\eqref{BiBa} 
\begin{equation}\label{vecident}
\frac{d}{dR}E^{V}_M + E^{V}_M + \frac{(D-3)}{\mK} \nabla^N R^T_{NM}=0 
\end{equation}
asserts that the vector constraint equation is automatically solved at 
all values of $R$ if its solved at one value of $R$ (we use here that 
we have already solved the tensor equation so that $R^T_{AB}=0$).

We will find it convenient to solve the vector constraint equation at 
$R=0$. From Table \ref{Ebasis} we see that 
$$E^V_M=\frac{1}{2}\frac{\mK}{(D-3)}(1-e^{-R}) \frac{d}{dR} \left(  \frac{\nabla^M H^{(T)}_{MN}}{(D-3)} \right) + {\mathcal V}_M^{V}(R)$$
At $R=0$ 
$$E^V_M= {\mathcal V}_M^{V}(0)$$
It follows that the constraint equation is solved at $R=0$ if and only if 
$ {\mathcal V}_M^{V}(0)$ vanishes (here we use the fact that 
$H^{(T)}_{MN}$ is regular at $R=0$; see the previous subsection) . 
This requirement is a statement of the membrane equations of motion. 

We would like to reemphasize that the membrane equations of motion at $n^{th}$ order 
are obtained simply by evaluating the $n^{th}$ order vector constraint equation at 
$R=0$. At $R=0$ this equation is independent of all the unknown $n^{th}$ order 
fluctuation fields. As a consequence the membrane equations of motion may be obtained 
at $n^{th}$ order {\it before} solving for the fluctuation fields at $n^{th}$ order, 
as in studies of the fluid gravity correspondence. 

The analysis presented in this subsection so far has been valid at 
every order in perturbation theory. Specializing now to the first order, we
read off the value of ${\mathcal V}_M^{V}(0)$ from Table \ref{Esrc}. Equating
this expression to zero we find the first order membrane equation of motion
 \begin{equation} \label{VE1}
\left(\frac{\nabla^2 u_{A}}{\mathcal{K}}- \frac{\nabla_A\mathcal{K}}{\mathcal{K}}+ u_{C}K^C_A-u.\nabla u_A \right)\mathcal{P}^A_B=0
\end{equation}
While all fields in \eqref{VE1} live in the full bulk spacetime $R^{D-1,1}$, 
and all derivatives in that equation are bulk spacetime derivatives, the 
equation \eqref{VE1} itself holds only on the membrane surface $\psi=1$. 
Using the subsidiary conditions \eqref{auxeq} it is possible to rewrite 
\eqref{VE1} as an equation restricted to the membrane. As demonstrated 
in \cite{Bhattacharyya:2015fdk} the equation of motion of motion turns 
out to take exactly the same form as \eqref{VE1} in this language. In other 
words \eqref{VE1} also holds true if we think of $K_{MN}$ and $u_M$ as 
membrane world volume fields, and regard every derivative in that equation 
as a covariant derivative on the membrane world volume.

\subsubsection{Solving for the vector fluctuation}

As we have explained in the previous subsubsection, the 
constraint vector equation is automatically solved at every 
$R$ provided the membrane equation is obeyed. Assuming this is 
the case, we have already solved one of the two vector equations. 

In order to solve for the unknown function,  $H^{(V)}_M$, in the vector sector, 
we now turn to the second vector equation $R_L^{V_2}=0$. This equation 
takes the form
\begin{equation} \label{dynveceq}
\left( \frac{-\mK^2}{2(D-3)^2} \right) e^{-2R}(-1+ e^{R}) \frac{d}{dR}(e^{R} \frac{d}{dR} H^{(V)}_{M}) + {\mS}^{V_2}_{M}(R) = 0
\end{equation}
As in the previous subsection we will proceed to solve \eqref{dynveceq} 
for an arbitrary source function, plugging in the first order result for 
the source 
\begin{equation}\label{fos}
{\mS}^{V_2,1}_{A}(R) = -\frac{\mK}{2(D-3)} e^{-2R}(-1+ e^{R}) 
\left( u^M K_{MN} - u^{M}\nabla_{M} u_{N} \right) \CP^N_A 
\end{equation}
only at the end of the computation.

Notice that the LHS of \eqref{dynveceq} vanishes at $R=0$. It follows 
that \eqref{dynveceq} admits regular solutions if and only if 
${\mathcal S}^{V_2}_M(R)$ also vanishes at $R=0$. It would naively seem that 
this requirement imposes a new constraint on membrane data, independent 
of \eqref{VE1}. \footnote{Had this step of the programme imposed a new 
constraint, we would have obtained a new membrane equation - and so obtained
more membrane equations than membrane variables, leading to an inconsistent
dynamical system.} However it turns out that the vanishing of 
${\mathcal S}^{V_2}_M(R)$ is automatic; indeed it follows from \eqref{DCrel} that 
$R^{V_2}_M$ is simply identical to the vector constraint equation 
$E^{V}_M$ at $R=0$. It follows as a consequence that ${\mathcal S}^{V_2}_M(R)$ 
is proportional to the LHS of  
\eqref{VE1} at $R=0$. \footnote{ To see this we note that \eqref{dynveceq} 
reduces to ${\mathcal S}^{V_2}_M(R)$ at $R=0$ while 
$E^{V}_M$ reduces to the LHS of \eqref{VE1} at $R=0$.}. 

Using the fact that ${\mS}^{V_2,1}_{M}(0)$ vanishes, we integrate
\eqref{dynveceq} once to find 
\begin{equation}
\begin{split}
& e^{R} \frac{d}{dR} H^{(V)}_{M} = \left( \frac{-2(D-3)^2}{\mK^2} \right) 
\left[ \int_0^R  \left( \frac{-e^{y}}{1-e^{-y}} \right){\mS}^{V_2}_{M}(y)dy 
+ C_M^{V_2} \right]
\end{split}
\end{equation}
where $C_M^{V_2}$ is an as yet undetermined integration constant. Integrating 
a second time we find 
\begin{equation}\label{vecsolf}
 \begin{split}
&   H^{(V)}_{M} = \left( \frac{2(D-3)^2}{\mK^2} \right) \int_R^{\infty} e^{-x}  \left[ \int_0^x  \left( \frac{-e^{y}}{1-e^{-y}} \right){\mS}^{V_2}_{M}(y)dy  \right]dx 
- C_M^{V_2} e^{-R}
\end{split}
\end{equation}
The upper limit on the the outer integral of \eqref{vecsolf} has been 
determined from the requirement that $H^{(V)}_{M}$ vanishes at large $R$. 
The expression for $H^{V}_M$ may be simplified by integrating by parts; we 
find 
\begin{equation}\label{vecsolsu}
H^{(V)}_{M}(R) = \left( \frac{2(D-3)^2}{\mK^2} \right)
\left( e^{-R}  \int_0^R  \left( \frac{-e^{x}}{1-e^{-x}} \right){\mS}^{V_2}_{M}(x)dx 
- \int_R^\infty \frac{ {\mS}^{V_2}_{M}(x)}{1-e^{-x}} \right)
- C_M^{V_2} e^{-R}
\end{equation}
In particular that 
\begin{equation}\label{vecsolc}
H^{(V)}_{M}(0) = -\left( \frac{2(D-3)^2}{\mK^2} \right)
 \int_0^\infty \frac{ {\mS}^{V_2}_{M}(x)}{1-e^{-x}} 
- C_M^{V_2} 
\end{equation}
It follows (see \eqref{velhor}) that 
\begin{equation}\label{vecint}
 C_M^{V_2} = -\left( \frac{2(D-3)^2}{\mK^2} \right)
 \int_0^\infty \frac{ {\mS}^{V_2}_{M}(x)}{1-e^{-x}} 
\end{equation}
so that 
\begin{equation}\label{vecsols}
H^{(V)}_{M}(R) = \left( \frac{2(D-3)^2}{\mK^2} \right)
\left( e^{-R}  \int_0^R  \left( \frac{-e^{x}}{1-e^{-x}} \right){\mS}^{V_2}_{M}(x)dx 
- \int_R^\infty \frac{ {\mS}^{V_2}_{M}(x)}{1-e^{-x}}
+e^{-R} \int_0^\infty \frac{ {\mS}^{V_2}_{M}(x)}{1-e^{-x}}
 \right)
\end{equation}

The expression \eqref{vecsols} is our final expression for $H^{(V)}_{M}(R)$
at any order in perturbation theory in terms of the source function at that 
order. Note that $H^{(V)}_{M}(R)$ is uniquely determined in terms of its 
source function; the integration constants in \eqref{dynveceq} are 
uniquely determined by the requirement that $H^{(V)}_{M}(R)$ vanish at infinity 
and that \eqref{velhor} is obeyed at $R=0$. 

Plugging the first order expression for the source \eqref{fos} into 
\eqref{vecsols}, at first order we find 
\begin{equation}
 H^{(V,1)}_{M} = \frac{(D-3)}{\mathcal{K}}Re^{-R}\left( u^{A} K_{AN}-u^{A}\nabla_A u_N \right) P^N_M
\end{equation}
 
 \subsection{Scalar sector}

In the scalar sector we have four equations for the two variables $H^{(Tr)}$ and 
$H^{(S)}$. As a basis for the four equations we find it convenient to use the 
two scalar constraint equations $E^{S_1}$ and $E^{S_2}$ (see Table \ref{Ebasis})
together with the two additional equations $R^{S_1}=0$ and $R^{S_2}=0$ 
(see Table \ref{Rbasis}).

\subsubsection{Constraint Equations and $\nabla.u$}
As in the previous subsection it is consistent to have four equations 
for two variables only if two of the four equations are identities. The last two 
equations in \eqref{BiBa} 
\begin{equation}\label{scaconst}
\begin{split} 
&\frac{d}{dR}E^{S_1} + E^{S_1} + \frac{(D-3)}{\mK} \nabla^N R^{V_2}_{N}=0 \\ 
 &\frac{d}{dR}E^{S_2} + (1-\frac{1}{2}e^{-R})R^{S_1} + R^{S_2} + \frac{(D-3)}{\mK}\frac{1}{1-e^{-R}} \nabla^M \left(  E_M^V- R^{V_2}_{M} \right) =0
\end{split}
\end{equation}
assert that this is indeed the case. As we have already solved the vector sector at $n^{th}$ order 
$R^{V_2}_{N}$ vanishes. It follows that the first equation in \eqref{scaconst} asserts that if 
$E^{S_1}$ is solved at any $R$ it is automatically solved at every $R$. 
When evaluated at $R=0$ this equation reduces to the condition 
\begin{equation}\label{condo}
{\cal V}^{S_1}(0)+ \frac{\mK}{2(D-3)}~\nabla.u =0
\end{equation}
Recall that at leading order $\nabla.u=0$. \eqref{condo} determines the 
correction to this statement at subleading orders. 

As in the previous subsection we emphasize that the expression for $\nabla.u$ at 
$n^{th}$ order is determined simply by evaluating the $n^{th}$ order 
constraint equation $E^{S_1}$ at $R=0$. In order to obtain this correction we 
do not need to solve for any of the $n^{th}$ order fluctuation fields, 
all of which drop out in $E^{S_1}$ evaluated at $R=0$.

The analysis of this subsection has, so far, been valid at every order in perturbation theory. 
Specializing to  first order it is easily verified from Table \ref{Esrc} that ${\mathcal V}^{S_1}(0)=0$.
It follows that the zero order relation $\nabla.u=0$ is uncorrected at 
first order~~(since $(\nabla . u)_{0}=\mathcal{V}^{S_1}(0)=0 $). As we will see in the next section, the situation is 
different at second order. 

The constraint equation $E^{S_2}$ plays a distinct logical role from $E^{S_1}$ in our perturbative
programme. Once the tensor and vector equations had been solved, \eqref{scaconst} assured us 
that $E^{S_1}(R)$ obeys a homogeneous differential equation in $R$ (see 
\eqref{bianchirel} which makes no reference 
to any of the other equations in the scalar sector. On the other hand the differential 
equation obeyed by $E^{S_2}$ involves the other scalar equations (see the last equation in 
\eqref{biarel}). The most useful way to view the last equation in \eqref{biarel} is as 
follows. It might, a priori, have seemed that we have 4 equations in the scalar sector. 
We have already dealt with $E^{S_1}$ above leaving behind a three dimensional space of 
equations. A useful basis for this space is given by $E^{S_2}$, $R^{S_1}$ and $R^{S_2}$.
The last equation in \eqref{biarel} allows us to eliminate $R^{S_2}$ from this basis. 
In order to complete solving in the scalar sector we need only solve the equations 
$E^{S_2}$, $R^{S_1}$. In other words the constraint
equation $E^{S_2}$ does not constrain data: instead it may be used to solve for the 
scalar fluctuation. We turn to this task in the next subsubsection.

\subsubsection{Solving for the scalar fluctuations}

The equation $R^{S_1}$
 \begin{equation}
 R^{S_1}=\left( \frac{-\mK^2}{2(D-3)^2} \right) \frac{d^2 H^{(Tr)}}{dR^2}+{\mS}^{S_1}(R)=0
 \end{equation}
is easily solved. Integrating the above equation once we get
 \begin{equation}
   \frac{d H^{(Tr)}}{dR} = \left(\frac{-2(D-3)^2}{\mK^2}\right) \int_R^{\infty} dx~{\mS}^{S_1}(x)
 \end{equation}
Where we have fixed the boundary condition from the requirement that  $H^{(Tr)}$ and so its 
derivative $\frac{d H^{(Tr)}}{dR}=0$ vanish at large $R$. 
Integrating this equation once again we have
\begin{equation}\label{htreq}
\begin{split}
  H^{(Tr)} &= \left(\frac{2(D-3)^2}{\mK^2}\right) \int_R^{\infty} dy \int_y^{\infty} dx~ {\mS}^{S_1}(x) \\
  &= \left(\frac{2(D-3)^2}{\mK^2}\right) \left[ -R \int_R^{\infty} dx~ {\mS}^{S_1}(x) + \int_R^{\infty}dx~ x~{\mS}^{S_1}(x) \right] 
\end{split}
 \end{equation}
where, once again we have fixed the integration constant from the requirement that 
$H^{(Tr)}=0$ at large $R$.

Specializing now to first order we note ${\mS}^{S_1,1}=0$ so that  
\begin{equation}
 H^{(Tr,1)}=0
\end{equation}

The equation $E^{S_2}$ takes the form 
\begin{equation}\label{sctt} 
\begin{split}
  &\frac{d}{dR} (H^{(S)}e^{R}) =\frac{2(D-3)^2}{\mK^2}e^{R}\mathcal{S}_{S}(R) ~~~~\text{where,} \\
  &\mathcal{S}_{S}(R)= -\frac{\mathcal K}{2 (D-3)} \frac{d}{dR}\left( \nabla^M H^{(V)}_M \right)-\frac{\mK}{(D-3)} \nabla^M H^{(V)}_M \\
  &+\frac{\mK^2}{4(D-3)^2}(2-e^{-R})\frac{d}{dR}H^{(Tr)}  - \frac{1}{2} \nabla^M\nabla^N H^{(T)}_{MN} + {\mathcal V}^{S_2}(R) + \frac{\mK}{2(D-3)}e^{-R}~\nabla.u
\end{split}
 \end{equation}

Plugging in the already obtained 
expressions of $H^{(V)}_M ~~,H^{(T)}_{MN}~, ~H^{(Tr)} $ (see \eqref{vecsols},\eqref{htreq} and \eqref{si}) and using \eqref{BiBaSrc}, the source function 
${\mathcal S}_S(R)$ can be rewritten as a linear functional of the elementary sources 
${\mathcal S}^{S_1}$, ${\mathcal S}^{S_2}$ and ${\mathcal V}^{S_1}$ 
\footnote{It turns out that all dependence on the fourth independent scalar 
source, ${\mathcal V}^{S_2}$ cancels.}. Upon simplifying (by integrating 
by parts on several occasions) we find 
\begin{equation}\label{compss}
\begin{split}
 \mathcal{S}_{S}(R)&=  \int_{R}^{\infty}{\mS}^{S_2}(x)dx +\frac{1}{2}\int_{R}^{\infty}(2-e^{-x}){\mS}^{S_1}(x)dx -\frac{1}{2}(2-e^{-R})\int_{R}^{\infty}{\mS}^{S_1}(x)dx\\&-\left(1-e^{-R}\right)\int_{R}^{\infty}\left(\frac{e^{x}\left({{\mathcal V}^{S_1}}^{'}(x)+{\mathcal V}^{S_1}(x)\right)}{(e^{x}-1)}dx\right)dy-\mathcal{V}^{S_{1}}(R)+e^{-R}\mathcal{V}^{S_1}(0)\\&+\log (1-e^{-R})\left({{\mathcal V}^{S_1}}^{'}(0)+{\mathcal V}^{S_1}(0)\right) + (\nabla \cdot u)\frac{\mK e^{-R}}{2(D-3)}
 \end{split}
 \end{equation}

We note that ${\mathcal S}_S$ is analytic at $R=0$ if and only if 
\begin{equation} \label{sccond}
  {{\mathcal V}^{S_1}}^{'}(0)+{\mathcal V}^{S_1}(0)=0 
 \end{equation}
This condition is, in fact, automatic. It follows from the second of 
\eqref{BiBaSrc} that the LHS of \eqref{sccond} is proportional to 
$\nabla^N{\mathcal S}_N^{V_2}(0)$. We have already argued, however, that 
${\mathcal S}_N^{V_2}$ vanishes at $R=0$. Since this condition holds at 
every point on the membrane, it follows also that $\nabla^N{\mathcal S}_N^{V_2}(0)
=0$
establishing \eqref{sccond}. 
\footnote{In studies of the fluid gravity correspondence a derivative 
of the equation of the $n^{th}$ order equation contributes to sources 
only at $(n+1)^{th}$ order in the derivative expansion. In the large 
$D$ expansion of this paper, however, the suppression in order resulting 
from using an extra derivative can be compensated for by an enhancement 
in order resulting from the contraction of a spacetime index. Consequently
the equation of motion and its contracted derivatives are of the same 
order in the large $D$ expansion.}

Plugging \eqref{compss} into \eqref{sctt}, integrating (and simplifying 
using integration by parts) we find 
\begin{equation}\label{hsans}
\begin{split}
 H_{S}(R)&=\frac{2(D-3)^2}{\mK^2}e^{-R}\Bigg(\frac{(\mK(\nabla \cdot u))R}{2(D-3)} +  e^{R}\int_{R}^{\infty}{\mS}^{S_2}(x)dx -\int_{0}^{\infty} {\mS}^{S_2}(x)dx+\int_{0}^{R}e^{x}{\mS}^{S_2}(x)dx  \\&+ \frac{e^R}{2}\int_{R}^{\infty}(2-e^{-x}){\mS}^{S_1}(x)dx+\frac{1}{2}\int_{0}^{R}e^x(2-e^{-x}){\mS}^{S_1}(x)dx-\frac{1}{2}\int_{0}^{\infty}(2-e^{-x}){\mS}^{S_1}(x)dx\\&-\frac{1}{2}(2 e^R-R)\int_{R}^{\infty}{\mS}^{S_1}(x)dx+\int_{0}^{\infty}{\mS}^{S_1}(x)dx-\frac{1}{2}\int_{0}^{R}(2e^y-y){\mS}^{S_1}(x)dx\\&-\int_{0}^{R}\left(e^{y}-1\right)\int_{y}^{\infty}\left(\frac{e^{x}\left({{\mathcal V}^{S_1}}^{'}(x)+{\mathcal V}^{S_1}(x)\right)}{(e^{x}-1)}dx\right)dy-\int_{0}^{R}e^x{\mathcal V}^{S_1}(x)dx+R \mathcal{V}^{S_1}(0)\Bigg) 
 \end{split}
 \end{equation}
Explicitly at first order
 \begin{equation}\label{hsfirst}
  H^{(S,1)} = \frac{D-3}{\mathcal{K}}R e^{-R}\left( R \left( -\frac{\mathcal{K}}{D-3}-\frac{u\cdot\nabla \mathcal{K}}{\mathcal{K}}+\frac{u\cdot K \cdot u}{2} \right) +\left( \frac{\mathcal{K}}{D-3}+u\cdot K \cdot u \right) \right)
 \end{equation}

 \subsection{Final Result for the first order metric}
After integrating the ordinary differential equations corresponding to Einstein's equations and imposing the condition that the metric is regular at the horizon, matches flat space at the end of the membrane region and \eqref{velhor}, we get the following solutions for the various components of the metric correction.
\begin{equation}
\begin{split}
& H^{(T,1)}_{MN} = 0 \\
& H^{(Tr,1)} = 0 \\
& H^{(V,1)}_{M} = \frac{(D-3)}{\mathcal{K}}Re^{-R}\left( u^{A} K_{AL}-u^{A}\nabla_A u_L \right)\CP^L_M\\
& H^{(S,1)} = \frac{D-3}{\mathcal{K}}R e^{-R}\left( R \left( -\frac{\mathcal{K}}{D-3}-\frac{u\cdot\nabla \mathcal{K}}{\mathcal{K}}+\frac{u\cdot K \cdot u}{2} \right) +\left( \frac{\mathcal{K}}{D-3}+u\cdot K \cdot u \right) \right)
\end{split}
\end{equation}
Thus we can write the 1st order corrected metric as
\begin{equation}\label{met1}
\begin{split}
&g_{MN} = \eta_{MN} +\frac{O_M O_N}{\psi^{D-3}}\\ &+ \frac{1}{D-3}\bigg[ \frac{D-3}{\mathcal{K}}R e^{-R}\left( R \left( -\frac{\mathcal{K}}{D-3}-\frac{u\cdot\nabla \mathcal{K}}{\mathcal{K}}+\frac{u\cdot K \cdot u}{2} \right) +\left( \frac{\mathcal{K}}{D-3}+u\cdot K \cdot u \right) \right)O_MO_N\\ &+ \frac{(D-3)}{\mathcal{K}}Re^{-R} \left( u^{A} K_{AL}-u^{A}\nabla_A u_L \right) P^L_{(M} O_{N)} \bigg]
\end{split}
\end{equation}

\section{2nd order solution}

The metric \eqref{met1} solves Einstein equation to first subleading order. In this section we implement the perturbative procedure 
to one higher order. In other words we determine the correction $H^{(2)}_{MN}$ in a way that 
ensures that $R_{AB}$ evaluated on the corrected metric is of order $1/D$ (more precisely that 
$R_{AB}R^{AB}$ is of order $1/D^2$). 

The procedure we follow is exactly that of the previous section: in fact second order corrections to the 
metric are given directly by the formulae of the previous subsection with one modification: we need to use the second order rather than first order source functions. In other words the computation at second order 
boils down entirely to determining the second order sources. 

In order to determine the sources at second order we plug the first order corrected metric \eqref{met1} 
together with an as yet undetermined second order correction $h^2_{MN}$ 
into Einstein's equations. We use the fact that the shape and velocity functions in the first order 
corrected metric obey the equation of motion
\begin{equation}\label{meom}
 \left(\frac{\nabla^2 u_{A}}{\mathcal{K}}- \frac{\nabla_A\mathcal{K}}{\mathcal{K}}+ u_{C}K^C_A-u.\nabla u_A \right)\mathcal{P}^A_B + \frac{1}{D}\mathcal{E}_A \CP^A_B =0
\end{equation}
where $\mathcal{E}_B$ is an as yet undetermined `2nd order' correction to the equations of motion. 
As in the previous subsection we solve the equations in the neighbourhood of a particular point on the 
event horizon. In our analysis, however, we use the fact that the membrane equations of motion 
\eqref{meom} are obeyed not just at the particular point we are expanding about but everywhere on the 
membrane. In other words we use the fact that the derivative of \eqref{meom} vanishes at the point of 
interest. Finally we also use the fact that $\nabla.u$ is an as yet undetermined quantity of order 
$1/D$.

We find by explicit computation that the curvature components listed table \ref{Rbasis} do indeed take the 
form listed in table \ref{Rbasisex},\ref{Ebasis} once all metric fluctuation fields in that table are identified with second order 
fluctuations. Our explicit computations also yield explicit expressions for all the second order source functions. We present an explicit listing of these source functions in Tables \ref{Rbsrc2} and \ref{Esrc2} 
in the Appendix. 

In the rest of this section we obtain the second order correction to the metric by inserting the 
second order sources listed above into the general integral formulae of the previous section and 
performing all integrals.

\subsection{Constraints on membrane data}

\subsubsection{Correction to the membrane equations from the vector 
sector}

As in the previous subsection \eqref{vecident} guarantees that  the 
vector constraint equation $E^V_M=0$
is solved at any $R$ if the equation is obeyed at $R=0$. As in the previous subsection the constraint equation at $R=0$ is independent of the second order fluctuation fields. From table \ref{Esrc2} we see that this constraint 
equation at $R=0$ determines $-\frac{1}{D}\mathcal{E}_A \CP^A_B$ - the second order correction
to the membrane equation of motion - in terms of appropriate expressions involving the 
membrane extrinsic curvature and velocity fields. Adding these correction terms to the 
first order membrane equation \eqref{VE1copy} we recover the second order corrected membrane 
equation

\begin{eqnarray}\label{veq1geom}
\nonumber&&\Bigg[\frac{\nabla^{2}u}{\mathcal{K}}-\frac{\nabla \mathcal{K}}{\mathcal{K}}+u\cdot K-(u\cdot\nabla )u \bigg]\cdot \CP +\bigg[\frac{\nabla^2\nabla^2 u}{\mathcal{K}^3}-\frac{\nabla (\nabla^2 \mathcal{K})}{\mathcal{K}^3} \\\nonumber&+&3\frac{(u\cdot K\cdot u)(u\cdot \nabla u)}{\mathcal{K}}-3\frac{(u\cdot K\cdot u)(u\cdot \nabla n)}{\mathcal{K}}-6\frac{(u \cdot (\nabla^2 n))(u\cdot \nabla u)}{\mathcal{K}^2}\\&+&6\frac{(u \cdot (\nabla^2 n))(u\cdot \nabla n)}{\mathcal{K}^2}+\frac{3}{D-3} u \cdot \nabla u-\frac{3}{D-3}u \cdot \nabla n\Bigg]\cdot \CP=0
\end{eqnarray}
where
\begin{equation}
\CP^{AB}= \eta^{AB}- n^A n^B +u^A u^B
\end{equation}
The 1st square bracket in \eqref{veq1geom} is simply the 1st order equation of motion while the 2nd square bracket represents subleading corrections. \footnote{Note that we can write the equation \eqref{veq1geom} in a nicer looking form by using the subsidiary conditions \eqref{auxeq}, divergence of first order membrane equation of motion \eqref{VE1copy} and divergence of velocity condition \eqref{Divu}. The form is
\begin{equation}
 \left( \frac{\nabla^2O}{\nabla.O}+O.\nabla O \right)\cdot \CP + \left( \frac{\nabla^2\nabla^2 O}{(\nabla.O)^3} + 3\frac{\nabla^2(\nabla.O)}{(\nabla.O)^3}O.\nabla O \right)\cdot \CP = 0
\end{equation}
}

We would like, however, to emphasize an important technical 
point. All the fields in \eqref{veq1geom} are assumed to live in all of the embedding flat spacetime; 
they are extended off the surface of the membrane by the subsidiary conditions listed earlier in this paper. 
While all covariant derivatives listed in \eqref{veq1geom} are evaluated on the surface of the membrane, 
they act on fields defined in all of spacetime. 

As the membrane equations of motion are intrinsic to the membrane, it is clearly unnatural 
to write them in terms of spacetime derivatives of an essentially arbitrary extension of 
membrane fields into the embedding spacetime. The equation of motion \eqref{veq1geom} can be 
rewritten so that all fields in that equation are purely membrane world volume fields, and every 
derivative in the equation is a covariant derivative on the membrane world volume. We now 
explain how this is done. 

The relationship between the bulk covariant derivatives 
of tensors (e.g. $u_M$) and membrane worldvolume derivatives of the same quantities is quite 
straightforward when no more than one derivative acts on the same object. The spacetime covariant 
derivative is obtained from the corresponding bulk quantity by projecting every index (not just the 
derivative indices) onto the membrane world volume. However this relationship is more complicated
when we have two or more derivatives acting on the same object; the reason for the additional complication
is that the formula for multiple worldvolume covariant derivatives involves inserting projectors at each 
step (when you define the first derivative in terms of bulk derivatives, then again when you define 
the second derivative in terms of bulk derivatives etc); when such expressions are opened out, outer 
derivatives act on projectors used to define the inner derivatives. Tracing through 
the required algebra we find that the corrected second order membrane equation of motion, 
written in terms of fields and covariant derivatives that live purely on the membrane 
world volume, takes the form
\begin{eqnarray}\label{veq1geomwv}
\nonumber&&\Bigg[\frac{\nabla^{2}u_{A}}{\mathcal{K}}-\frac{\nabla_{A} \mathcal{K}}{\mathcal{K}}+u^{B} K_{BA}-u\cdot\nabla u_{A}\Bigg]\CP^{A}_{C} \\\nonumber&+& \Bigg[\left(-\frac{u^{C}K_{CB}K^{B}_{A}}{\mathcal{K}}\right)+\left(\frac{\nabla^2\nabla^2 u_{A}}{\mathcal{K}^3}-\frac{u \cdot \nabla \mathcal{K}\nabla_{A}\mathcal{K}}{\mathcal{K}^3}-\frac{\nabla^{B}\mathcal{K}\nabla_{B}u_{A}}{\mathcal{K}^2}-2\frac{K^{CD}\nabla_{C}\nabla_{D}u_{A}}{\mathcal{K}^2}\right) \\ \nonumber &+&\left(-\frac{\nabla_{A}\nabla^2 \mathcal{K}}{\mathcal{K}^3} +\frac{\nabla_{A}\left(K_{BC}K^{BC}\mathcal{K}\right)}{\mathcal{K}^3}\right) + 3\frac{(u\cdot K\cdot u)(u\cdot \nabla u_{A})}{\mathcal{K}}-3\frac{(u\cdot K\cdot u)(u^{B} K_{BA})}{\mathcal{K}}\\  &-&6\frac{(u \cdot \nabla \mathcal{K})(u\cdot \nabla u_{A})}{\mathcal{K}^2}+6\frac{(u \cdot \nabla \mathcal{K})(u^{B}K_{BA})}{\mathcal{K}^2} + \frac{3}{D-3} u \cdot \nabla u_{A}-\frac{3}{D-3} u^{B}K_{BA} \Bigg]\CP^{A}_{C}=0\nonumber\\
\end{eqnarray}
The projector $\CP^{AB}$ used in this equation
\begin{equation}
\CP^{AB}= \tilde{g}^{AB} + u^A u^B
\end{equation}
where $\tilde{g}^{AB}$ is the induced metric on the world volume of the membrane. 

The equation \eqref{veq1geomwv} can be slightly simplified as follows. 
Let us first note that \eqref{veq1geomwv} takes the schematic form 
\begin{equation}\label{seomo}
F^A + \frac{S^A}{ \mK}=0
\end{equation}
where $F^A$ is the first order contribution to the equation of motion (the first line 
of \eqref{veq1geomwv}) while $\frac{S^A}{\mK}$ is the second order contribution (the second-fourth lines of \eqref{veq1geomwv}). $F^A$  and $S^A$ are each vector fields of order unity.

Let us now consider the modified equation of motion 
\begin{equation}\label{seomt}
F^A + \frac{S^A}{ \mK} + \nabla. F \frac{\zeta^A}{\mK^2}=0
\end{equation}
where $\zeta^A$ is any vector field of order unity. As $\nabla.F$ is naively 
of order $D$, the difference between the equations \eqref{seomt} and \eqref{seomo}
is naively of order $\frac{1}{D}$ suggesting that \eqref{seomo} and \eqref{seomt} 
differ at first subleading order. This is not the case. By taking a divergence of 
either \eqref{seomo} or \eqref{seomt}, the reader can easily convince herself that, 
onshell, $\nabla.F$ is of order unity (rather than the naive estimate of order $D$). 
If follows that \eqref{seomt} and \eqref{seomo} actually differ only at second 
subleading order ( $\frac{1}{D^2}$ ) and are equivalent at first subleading order. 
We are thus allowed to simplify \eqref{veq1geomwv} by adding any expression of the form 
$\nabla. F \frac{\zeta^A}{\mK^2}$ to it. 

Now it was demonstrated in \cite{Bhattacharyya:2015fdk} that 
\begin{equation}\label{SE1}
 \frac{\nabla.F}{\mK}=\frac{\nabla^2\mK}{\mK^2} - 2~\frac{u.\nabla\mK}{\mK} + u.K.u 
\end{equation}
Using this relation and making the the choice 
\begin{equation}\label{zcho}
\zeta^A= -3 \left( (u.\nabla u)_A - u_BK^B_A \right)
\end{equation}
we find that \eqref{veq1geomwv} is equivalent to \eqref{seomt} whose explicit form
is 
\begin{eqnarray}\label{VE2alt}
\nonumber&&\Bigg[\frac{\nabla^{2}u_{A}}{\mathcal{K}}-\frac{\nabla_{A} \mathcal{K}}{\mathcal{K}}+u^{B} K_{BA}-u\cdot\nabla u_{A}\Bigg]\CP^{A}_{C} \\\nonumber&+& \Bigg[\left(-\frac{u^{C}K_{CB}K^{B}_{A}}{\mathcal{K}}\right)+\left(\frac{\nabla^2\nabla^2 u_{A}}{\mathcal{K}^3}-\frac{u \cdot \nabla \mathcal{K}\nabla_{A}\mathcal{K}}{\mathcal{K}^3}-\frac{\nabla^{B}\mathcal{K}\nabla_{B}u_{A}}{\mathcal{K}^2}-2\frac{K^{CD}\nabla_{C}\nabla_{D}u_{A}}{\mathcal{K}^2}\right) \\ \nonumber &+&\left(-\frac{\nabla_{A}\nabla^2 \mathcal{K}}{\mathcal{K}^3} +\frac{\nabla_{A}\left(K_{BC}K^{BC}\mathcal{K}\right)}{\mathcal{K}^3}\right) - 3\frac{\nabla^2\mK~u\cdot \nabla u_{A}}{\mathcal{K}^3}+3\frac{\nabla^2\mK~u^{B} K_{BA}}{\mathcal{K}^3}\\  &+& \frac{3}{D-3} u \cdot \nabla u_{A}-\frac{3}{D-3} u^{B}K_{BA} \Bigg]\CP^{A}_{C}=0\nonumber\\
\end{eqnarray}

\subsubsection{Divergence of velocity from a scalar constraint}

As we have explained in the previous section, the Einstein constraint equation 
$E^{S_1}$ is satisfied at all $R$ if it is satisfied at $R=0$. As explained
in the previous subsection, the equation at $R=0$ simply asserts that 
$$\nabla.u_{2}=- \frac{2(D-3)}{{\mathcal K}}{\mathcal V}^{S_1}(0)$$
Reading off the value of ${\mathcal V}^{S_1}(0)$ from the table \ref{Esrc2} we find 
\begin{equation}
\begin{split}
 \nabla\cdot u=\frac{(\nabla.u)_2}{D-3}=\frac{1}{2\mK}\left(\nabla_{(A}u_{B)}\nabla_{(C}u_{D)}\CP^{BC}\CP^{AD}\right)
\end{split}
 \end{equation}

\subsection{Second order corrections to the metric}\label{2ndmet}

\subsubsection{Tensor Sector}

The metric correction in the tensor sector is given by \eqref{si} 
\begin{equation}
\begin{split}
 H^{(T)}_{AB} &=\left( \frac{-2(D-3)^2}{\mK^2} \right) \int_R^{\infty}\frac{dy}{e^y-1}\int_0^y e^{x}{\mS}^{T}_{AB}(x) dx \\
 &= \left( \frac{2(D-3)^2}{\mK^2} \right) \bigg[ \log(1-e^{-R})\int_0^R e^x S^{T}_{AB}(x)dx + \int_R^{\infty} \log(1-e^{-x})e^x S^{T}_{AB}(x) \bigg]
\end{split}
\end{equation}
where ${\mS}^{T}_{AB}$ is the second order source listed in  table \ref{Rbsrc2}.
All the integrals that appear in the final answer can easily be performed
analytically, but the final results (given in terms of polylogs) 
are not very illuminating; we prefer to leave our final result  
in terms of an explicit integral.

\subsubsection{Vector Sector}

The solution for $H^{(V)}_M(R)$ at second order is given by \eqref{vecsols} 
\begin{equation} \label{vfin}
H^{(V)}_{M}(R) = \left( \frac{2(D-3)^2}{\mK^2} \right)
\left( e^{-R}  \int_0^R  \left( \frac{-e^{x}}{1-e^{-x}} \right){\mS}^{V2}_{M}(x)dx 
- \int_R^\infty \frac{ {\mS}^{V2}_{M}(x)}{1-e^{-x}}
+ e^{-R} \int_0^\infty \frac{ {\mS}^{V2}_{M}(x)}{1-e^{-x}}
 \right)
\end{equation}
with all sources read off  at 2nd order from table \ref{Rbsrc2}. 
As in the tensor sector, all integrals that appear in \eqref{vfin} 
can be explicitly performed in terms of polylogs, but we find the 
expression \eqref{vfin} in terms of explicit integrals more illuminating.

\subsubsection{Scalar Sector}

Equation $R^{S_1}$ is decoupled equation for $H^{(Tr)}$. The integrated form is given by \eqref{htreq} which we write again

\begin{equation}
\begin{split}
  H^{(Tr)} &= \left(\frac{2(D-3)^2}{\mK^2}\right) \int_R^{\infty} dy \int_y^{\infty} dx~ {\mS}^{S_1}(x) \\
  &= \left(\frac{2(D-3)^2}{\mK^2}\right) \left[ -R \int_R^{\infty} dx~ {\mS}^{S_1}(x) + \int_R^{\infty}dx~ x~{\mS}^{S_1}(x) \right] 
\end{split}
 \end{equation}

 The source ${\mS}^{S_1}$ for 2nd order is given in table \ref{Rbsrc2}. Substituting this we get the final form of the metric correction 

\begin{equation}\label{tracesolution}
H^{(Tr,2)}=-\left(\frac{2(D-3)^2}{\mK^2}\right)e^{-R}(1+R) \left( \left( u \cdot K-u\cdot \nabla u \right)\cdot \CP \cdot \left( u\cdot K-u\cdot \nabla u \right) \right)
\end{equation}

In a similar manner the fluctuation ${H}^{S}$ can is given by 
\eqref{hsans} upon plugging in the explicit values of the second order sources 
from Tables \ref{Rbsrc2},\ref{Esrc2}.

\section{The spectrum of small fluctuations around a spherical membrane}

The simplest solution of the second order membrane equations of motion is a static spherical membrane 
dual to a Schwarzschild Black hole. 
In this section we compute the spectrum of small fluctuations about this solution. Our answers  agree perfectly with earlier results for the spectrum of light quasinormal modes obtained by direct gravitational 
analysis, in \cite{Emparan:2014aba}. We regard this detailed agreement as a nontrivial consistency check of the 
second order membrane equations of motion derived in this paper. 

The computation presented in this section is a straightforward extension of the first order 
computation presented in section 5 of \cite{Bhattacharyya:2015fdk}. We have kept the discussion 
of this section brief. We refer the reader to section 5 of \cite{Bhattacharyya:2015fdk} for a fuller discussion 
of the logic behind our computation.

We work in standard spherical polar coordinates (see Eq 5.1 of \cite{Bhattacharyya:2015fdk}). The static spherical 
membrane is given by 
\begin{equation} \label{Schwcon}
r=1,  ~~~u=-dt,
\end{equation}
We study the small fluctuations 
\begin{equation}\label{SchwPert}
\begin{split}
r &= 1 + \epsilon~\delta r(t,\theta),\\
u &= -dt + \epsilon~\delta u_{\mu}(t,\theta)dx^{\mu}.
\end{split}
\end{equation}
about this solution and work to linear order in $\epsilon$. As explained in \cite{Bhattacharyya:2015fdk}, to linear order 
the metric on membrane worldvolume is given by 
\begin{equation}\label{SchwLin}
ds^2= -dt^2 + \left(1+2 \epsilon \delta r \right)d\Omega_{D-2}^2~~.
\end{equation}
As in \cite{Bhattacharyya:2015fdk} we find it convenient to work with covariant derivatives with respect to the unperturbed 
spherical metric
\begin{equation}\label{IndFlat}
ds^2=-dt^2+ d\Omega_{D-2}^2~~,
\end{equation}
The derivatives appearing from now on are all with respect to metric \eqref{IndFlat}. 
We use the following notation for the laplacian with respect to this fixed metric
$$\overline{\nabla}^2=\nabla_\mu\nabla^\mu=-\partial_t^2+\nabla_a\nabla^a=-\partial_t^2+\nabla^2$$

\subsection{The divergence condition}

The RHS of \eqref{Divu} is quadratic in $\epsilon$, and so vanishes upon linearizing in $\epsilon$. 
At linear order, therefore, \eqref{Divu} reduces to $\nabla.u=0$ (where the divergence is taken along the dynamical 
membrane world volume). As explained in \cite{Bhattacharyya:2015fdk}, this equation can be rewritten as  
\begin{equation}\label{velco1}
\nabla_\mu \delta u^\mu = - (D-2) \partial_t \delta r,
\end{equation}
where, the covariant derivatives \eqref{velco1} are now taken w.r.t. the fixed metric \eqref{IndFlat}. 
$u^0$ deviates from unity only at quadratic order in $\epsilon$. For the linearized considerations of this 
section, therefore, the LHS of \eqref{velco1} is simply the spatial divergence of the velocity
\begin{equation}\label{velco2}
\nabla_a \delta u^a = - (D-2) \partial_t \delta r.
\end{equation}
As in \cite{Bhattacharyya:2015fdk}, \eqref{velco2} may be solved by separating $u$ into its 
gradient and curl parts, i.e. by setting
\begin{equation}\label{veldec1}
\delta u_a = \nabla_a \Phi + \delta v_a,
\end{equation} 
with 
\begin{equation}\label{veldec2}
\nabla\cdot \delta v=0.
\end{equation}
It follows from \eqref{velco2} that 
\begin{equation} \label{phicon}
\nabla^2 \Phi= -(D-2) \partial_t \delta r.
\end{equation}

\subsection{Linearized equation of motion}

In order to obtain the linearized membrane equations of motion we use Eq 5.7 of 
\cite{Bhattacharyya:2015fdk} together with 
\begin{eqnarray} \nonumber
\frac{u^{E}K_{EB}K_{a}^{B}}{\mathcal{K}}&=&-\epsilon\frac{(\nabla_{a}\p_{t}\delta r-\delta u_{a})}{D-2}\\\nonumber
\frac{\nabla ^2\nabla ^2  u_{a}}{\mK^3}&=&\epsilon\frac{\bar{\nabla} ^2 \bar{\nabla} ^2 \delta u_{a}+\bar{\nabla }^2 \nabla_{a}\p_{t}\delta r}{(D-2)^3} \\\nonumber
\frac{K^{CD}\nabla_C\nabla_D u_{a}}{\mK^2}&=&\epsilon\frac{\bar{\nabla} ^2 \delta u_{a}- \nabla_{a}\p_{t}\delta r}{(D-3)(D-2)}\\\nonumber
\frac{\nabla_{a}\nabla ^2 \mK}{\mK^3} &=&-\epsilon\frac{\nabla_{a} \bar{\nabla}^2(\bar{\nabla}^2 \delta r+\delta r(D-2))}{(D-2)^3}\\\nonumber
\frac{\nabla_{a}(K^{BC}K_{BC}\mK)}{\mK^3}&=&\epsilon\frac{3\nabla_{a}(-\bar{\nabla}^2\delta r-\delta r(D-2))}{(D-3)(D-2)}\\\nonumber
\end{eqnarray}

(the equations above are accurate only to linear order in $\epsilon$ and all covariant 
derivatives are taken with respect to \eqref{IndFlat}). The linearized membrane equation 
is given by 
\begin{equation}\label{LinEE}
\begin{split}
&\left[ \left(1+\frac{\overline{\nabla}^2}{D-2}\right)\delta u_a + \nabla_a\left(1+\frac{\overline{\nabla}^2}{D-2}\right)\delta r-\partial_t\nabla_a\delta r \left( 1-\frac{1}{D-2} \right) - \partial_t\delta u_a \right] \\
& +\bigg[ \frac{\nabla_a\partial_t \delta r-\delta u_a}{D-2}+\frac{\overline{\nabla}^2\overline{\nabla}^2 \delta u_a+\overline{\nabla}^2 \nabla_a \partial_t \delta r}{(D-2)^3}+ 2\frac{-\overline{\nabla}^2 \delta u_a+\nabla_a \partial_t \delta r}{(D-3)(D-2)}+\frac{\nabla_a \overline{\nabla}^2(\overline{\nabla}^2 \delta r+(D-2)\delta r)}{(D-2)^3} \\&+ 3\frac{\nabla_a(-\overline{\nabla}^2 \delta r-(D-2)\delta r)}{(D-3)(D-2)} + 3\frac{\partial_t \delta u_a}{(D-3)} + 3\frac{\partial_t \nabla_a \delta r-\delta u_a}{(D-3)} \bigg] =0.
\end{split}
\end{equation}
(\eqref{LinEE} generalizes equation (5.9) of \cite{Bhattacharyya:2015fdk}). 
 Substituting \eqref{veldec1} into \eqref{LinEE} we find the generalized version of of 
 (5.15) of \cite{Bhattacharyya:2015fdk}, 
\begin{equation} \label{VElin}
\begin{split}
&\left(\frac{\nabla^2}{D-2}+ 1  - \partial_t   +\frac{\overline{\nabla}^2 \overline{\nabla}^2  }{(D-2)^3}- \frac{2 (\nabla^2)}{(D-2)^2}+\frac{3 \partial_t}{(D-3)}-\frac{3  }{(D-3)}\right)\delta v_a=\\
& -\Bigg(\frac{\partial_t \nabla_a }{D-2}+ \frac{\nabla_a \nabla^2 }{D-2} + \nabla_a  -\nabla_a \partial_t  + \frac{2 \nabla_a \partial_t }{(D-2)^2}- \frac{\nabla_a \overline{\nabla}^2(\nabla^2  + (D-2))}{(D-2)^3}\\& - \frac{9 \nabla_a ( (D-2)^2 -(D-2)(9\nabla^2  - \partial_t^2 ))}{3(D-2)^3}+\frac{3 \partial_t \nabla_a }{(D-3)}\Bigg)\delta r\\
& -\Bigg(\frac{\nabla^2}{D-2}+ 1  - \partial_t   +\frac{\overline{\nabla}^2 \overline{\nabla}^2  }{(D-2)^3}- \frac{2 (\nabla^2)}{(D-2)^2}+\frac{3 \partial_t}{(D-3)}-\frac{3}{(D-3)}\Bigg) \nabla_a \Phi
\end{split}
\end{equation}

\subsection{Scalar quasinormal modes}

Using \eqref{velco2} and \eqref{phicon} we take the divergence of \eqref{VElin} to obtain
\begin{equation}\label{DivSca}
\begin{split}
&-(\overline{\nabla}^2+D-3)\partial_{t}\delta r+\frac{\partial_{t}\nabla^2 \delta r}{D-2}+\frac{\nabla^2\bar{\nabla}^2\delta r}{(D-2)}+\nabla^2 \delta r-\partial_{t}\nabla^2 \delta r-(D-2)\partial_{t} \delta r+(D-2)\partial_{t}^2 \delta r \\&+ \frac{\nabla^2\partial_{t} \delta r+ (D-2)\partial_{t} \delta r}{D-2} - \frac{ (\overline{\nabla}^2+D-3)^2(D-2)\partial_{t}\delta r+(\overline{\nabla}^2+D-3)\nabla^2\partial_{t}\delta r}{(D-2)^3} \\&+ 2\frac{(\overline{\nabla}^2+D-3)(D-2)\partial_t\delta r+\nabla^2\partial_{t}\delta r}{(D-2)^2}+\frac{\nabla^2\bar{\nabla}^2(\bar{\nabla}^2 \delta r  +\delta r(D-2))}{(D-2)^3}\\ &-\frac{\nabla^2(3\nabla^2 \delta r-\partial_t^2\delta r+3\delta r(D-2))}{(D-2)^2}-3\frac{D-2}{(D-3)}\partial_{t}^2 \delta r +\frac{3}{(D-3)}(\partial_{t} \nabla^2 \delta r +(D-2)\partial_{t} \delta r)=0
 \end{split}
\end{equation}
As in \cite{Bhattacharyya:2015fdk} we expand
\begin{equation}\label{ScaFlu}
\delta r= \sum_{l,m}a_{lm} Y_{lm} e^{-i \omega^r_l t}~~.
\end{equation} 
where the spherical harmonics $Y_{lm}$ obey 
\begin{equation} \label{ScalSph}
-\nabla_{S^{D-2}}^2 Y_{lm} = l(D+l-3) Y_{lm}.
\end{equation} 
Inserting \eqref{ScaFlu} into \eqref{DivSca} we obtain
\begin{equation}\label{ScaQNM}
\omega_l^r=\pm \sqrt{l-1}-i(l-1)+\frac{1}{D}\left(\pm \sqrt{l-1}\left(\frac{3l}{2}-2\right)-i(l-1)(l-2)\right)
\end{equation}
The result \eqref{ScaQNM} is in perfect agreement with the result obtained by EST in Equations (5.30) and (5.31) of \cite{Emparan:2014aba}.

As explained in \cite{Bhattacharyya:2015fdk}, the modes with $l=0$ and $l=1$ are special. At $l=0$ the formula 
\eqref{ScaQNM} yields  $\omega=0,2i-\frac{4i}{D}$. The second solution is, however, spurious
(see \cite{Bhattacharyya:2015fdk}). The first solution is the zero mode corresponding to rescaling the black hole; this 
is an exact zero mode at all orders in $1/D$.

At $l=1$ \eqref{ScaQNM} yields the frequencies $\omega=0,0$. As explained in \cite{Bhattacharyya:2015fdk} these two modes 
correspond to translations and boosts of the membrane.

\subsection{Vector quasinormal modes}

We expand the velocity fluctuations in a basis of vector spherical harmonic
\begin{equation}\label{VecFlu}
\delta v_a = \sum_{l,m} b_{lm} Y_a^{lm} e^{-i \omega^v_l t}
\end{equation}
Where, $l=1,2,3,...$. The vector spherical harmonics satisfy the property
\begin{equation} \label{VecSph}
\nabla^2 V=- [(D+l-3)l-1]V
\end{equation}

Plugging \eqref{VecFlu} into \eqref{VElin}, using \eqref{VecSph} and equating the coefficients of 
independent vector spherical harmonics (see \cite{Bhattacharyya:2015fdk} for more discussion) we obtain 
\begin{equation}\label{VecQNM}
\omega^v_l = -i(l-1)-\frac{i}{D} (l-1)^2.
\end{equation}

\eqref{VecQNM} is in perfect agreement with the formula (5.22) of \cite{Emparan:2014aba} derived earlier by EST  by purely 
gravitational analysis. Note that the mode with $l=1$ has vanishing frequency. As explained in 
\cite{Bhattacharyya:2015fdk} $l=1$ is the exact zero mode corresponding to setting the black hole spinning. 

\section{Discussion}

In this paper we have worked out the duality between the dynamics of black holes in a large number of dimensions
and the motion of a non gravitational membrane in flat space to second subleading order in $1/D$. Our work 
generalizes the analysis of \cite{Bhattacharyya:2015dva,Bhattacharyya:2015fdk}. The 
concrete new results of this paper are 
\begin{itemize}
 \item The second order corrected membrane equations of motion listed in \eqref{VE2copy}.
 \item The formula \eqref{Divu} for the divergence of the velocity field (which vanished at first order).
 \item The explicit form of the second order corrected metric dual to any given membrane motion 
 (see subsection \ref{2ndmet}
\end{itemize}

In addition to obtaining the new results listed above we have also achieved 
an improved understanding of the structure of the perturbative expansion in 
$1/D$. We have demonstrated that the perturbative programme,
implemented to first nontrivial order in 
\cite{Bhattacharyya:2015dva,Bhattacharyya:2015fdk}, can systematically be 
extended to every order in the $1/D$ expansion. In particular we have shown 
that the algebraically nontrivial `integrability' properties  that 
allowed for the existence of a first order solution in 
\cite{Bhattacharyya:2015dva,Bhattacharyya:2015fdk} are actually automatic at 
all orders as as a consequence of the well known equation \eqref{bianchi}. 

We have also explained that the membrane equations may directly be obtained 
by evaluating the Einstein constraint equation on the event horizon. 
In particular the membrane equations at $(n+1)^{th}$ order in $1/D$ are 
obtained by evaluating the constraint equations on $n^{th}$ order metric, 
without needing to solve for the $(n+1)^{th}$ order metric. We have also 
explained that the assumption of $SO(D-p-2)$ isometry, made in  
\cite{Bhattacharyya:2015fdk}, is not necessary; the membrane equations 
can be derived under much more general conditions

The fact our membrane equations arise from the Einstein constraint equations at the event horizon is 
strongly reminiscent of the `traditional' membrane paradigm of 
black hole physics. It would be very interesting to better understand the relationship between the the 
large $D$ membrane and the traditional membrane paradigm.
\cite{Price:1986yy,Price:1988sci,PhysRevD.18.3598}.

As black holes are thermodynamical objects, the black hole membrane studied in 
\cite{Bhattacharyya:2015dva,Bhattacharyya:2015fdk} and this paper should carry an entropy current. At leading order
in $1/D$ it turns out (see \cite{bh}) that this 
entropy current is given simply by a constant times $u^M$. The divergence of this entropy current is thus proportional to 
$\nabla.u$. It follows that the RHS of the formula \eqref{Divu} gives an expression 
for the rate of entropy production on the membrane. It would be interesting to further investigate this observation 
and its consequences.

On a related note, it would be interesting to derive the most general stationary solution of the second order 
corrected equations of motion derived in this paper and compare our results with those of \cite{Suzuki:2015iha}. 

In this paper we have focused  our attention on black holes propagating in an otherwise perfectly flat spacetime. 
It would be interesting to generalize our study to the motion of black holes propagating in any vacuum solution 
of Einstein's equations, e.g. a gravity wave. Such a generalization would allow us, for instance, to 
study the absorption of gravity waves by black holes at large $D$. At first order in the derivative 
expansion we expect the generalized effective membrane equation to be given simply by covariantizing first order flat
space equations of motion. At second order, however, the equations of motion could receive genuinely new 
contributions from the background Riemann tensor of the space in which the black hole propagates 
\footnote{Something similar happens in the study of forced fluids in the fluid gravity correspondence \cite{Bhattacharyya:2008ji}}. 
It would be interesting to work this out in detail. 

Finally, it would be interesting to put the membrane equations derived in this paper to practical use to allow us 
to learn new things about black holes. One possible direction would be to test out how well the large D expansion 
does in astrophysical contexts (i.e. when $D=4$). Another direction would be to use the formalism developed herein to 
address interesting unanswered structural questions about gravity, e.g. questions about the second law of 
thermodynamics in higher derivative gravity. We leave such investigations for the future.

\vskip .8cm 

\section*{Acknowledgments}
We would like to thank K. Inbasekar , S. Thakur and M. Mandlik for many useful 
discussions during the progress in the project. We would like especially to 
thank S. Bhattacharyya for several very useful discussions and explanations. 
S. Mazumdar would like to acknowledge the hospitality of the University of 
Barcelona , ETH Zurich and IISER Mohali while this work was in progress. 
Y.D. and S. Mazumdar would also like to thank ICTP,Trieste  for hospitality 
while this work was in progress. Y.D. ,A.D., S. Mazumdar and  A.S would 
also like to thank the organizers of The Fourth Indo-Israel Meeting, Goa 
for hospitality while this work was in progress . S.M. would like to 
thank IAS Princeton for hospitality while this work was in progress. The work
of all authors was supported by the Infosys Endowment for the study of the 
Quantum Structure of Spacetime, as well as an Indo Israel (UGC/ISF) grant. 
Finally we would all like to acknowledge our debt to the people of India 
for their steady and generous support to research in the basic sciences.

\appendix

\section{Method of calculation}

In this Appendix we outline the method we have employed to obtain the results quoted in 
tables \ref{Rbasisex}, \ref{Ebasis}, \ref{Rbsrc}, \ref{Esrc}, \ref{Rbsrc2}, \ref{Esrc2}.

As we have mentioned in the main text, our starting point is the metric listed in \eqref{pertthy},\eqref{Odef},\eqref{corrcon},\eqref{metcorr}. 
In order to obtain the equations of motion listed in table \ref{Rbasisex} (see also table \ref{Ebasis}) we simply 
plugged this metric into the vacuum Einstein equations. Assuming these equations are already obeyed at $n-1$ order 
we then obtained the form of the $n^{th}$ order equations. As emphasized in table \ref{Rbasisex}, each of these 
equations have a `homogeneous' contribution and a `source' contribution. The homogeneous contribution is 
linear in the (as yet unknown) $n^{th}$ order fluctuation, and takes the same form at all orders. In order to 
evaluate the homogeneous contribution to all equations of motion, consequently, it is sufficient to work at first 
order. 

While the first order computation is straightforward to perform analytically in principle, in practice the 
computations involved are rather lengthy \footnote{These computations have, however, also been performed analytically
in the upcoming paper \cite{bhattacharyya:2016ab}}. In order to guard against error we employed Mathematica in 
our computations using the following device. Following \cite{Bhattacharyya:2015dva,Bhattacharyya:2015fdk} we specialized to the particular case of 
metrics that preserve an $SO(D-p-2)$ isometry. Such special metrics effectively depend only on $p+3$ variables. 
For small values of $p$, therefore, all computations can be effectively performed on Mathematica (see 
\cite{Bhattacharyya:2015fdk} for a detailed explanation of how this is done). The first order computation 
performed in this manner yields the homogeneous part of the differential equations listed in tables 
\ref{Rbasisex} and \ref{Ebasis} in a straightforward manner. Note that the homogeneous part of
the equations are differential operators only in the variable $R$. They are `ultra-local' on the membrane. 
Consequently, even though the assumption of isometry was used as a trick to facilitate the computation 
of the homogeneous part of the equation, the final result obtained for the structure of the equations 
listed in tables \ref{Rbasisex} and \ref{Ebasis} is valid assuming only that all background 
quantities (e.g. ${\mathcal K}$) scale in the manner assumed in the text. In particular the 
homogeneous contribution to these equations are independent of $p$. By repeating all of our 
computations for $p=2$ and $p=3$ we have explicitly checked that this is the case. 

Apart from the homogeneous pieces, the equations listed in tables \ref{Rbasisex} and \ref{Ebasis} also have 
contributions from sources. Source terms are different at different orders in the computation. We obtained 
our explicit results for the first order sources listed in tables \ref{Rbsrc}, \ref{Esrc} and second order 
sources listed in tables \ref{Rbsrc2}, \ref{Esrc2} as follows. Working separately in the scalar, vector and 
tensor channels we first explicitly listed all possible source structures that could appear in any given 
equation both at first and second order in perturbation theory. The source structures that appear in our 
classification are the analogues of the 'geometrical' quantities listed in the LHS of Table 4 in \cite{Bhattacharyya:2015fdk}. At any 
given order, it follows that the sources that appear in the equations of tables \ref{Rbasisex} and \ref{Ebasis} 
are linear combinations of these structures with coefficients that are as yet unknown functions of $R$.
We then worked out the analogue of the RHS of Table 4 of \cite{Bhattacharyya:2015fdk}, i.e. we explicitly evaluated
each of these basis source terms in terms of `reduced source data' - the analogue of the expressions 
listed in table 1 of \cite{Bhattacharyya:2015fdk}. 

Using our explicit computations on Mathematica we read off the coefficients of all reduced sources in all of the 
equations listed in table \ref{Rbasisex} and \ref{Ebasis}. We then used our reduction formulae for `geometrical sources 
in terms of reduced sources' (analogue of Table 4 in \cite{Bhattacharyya:2015fdk}) to determine the coefficients of all source 
terms in the original geometrical basis of possible source terms. The last step (determination of geometrical 
sources from the known coefficients of reduced sources) is unambiguous provided the map between geometrical 
and reduced sources in invertible, i.e. provided there does not exist a nontrivial linear combination
of geometrical sources that maps to zero when re expressed in terms of reduced sources 
(i.e. vanishes under the the assumption of isometry). We have verified that this condition 
is met at first order provided $p \geq 2$ and at second order provided that $p \geq 3$. \footnote{It is 
easy to understand the inequalities listed here. When $p=1$, for instance, a potential source 
term proportional to the shear of the velocity field trivially vanishes just because fluids in one 
spatial dimension do not have a transverse direction in which to shear.}. This is the reason we 
performed our computations at $p=3$. \footnote{ We also performed all computations in $p=2$ and verified that 
we obtained the same results for all sources from this computation - except in the case of 
a single second order source that vanished at $p=2$ but not at $p=3$. The coefficient of this term was left 
undetermined at $p=2$ but we determined at $p=3$. }

\section{Sources at second order}

In this Appendix we present an explicit listing of all the sources that appear in the second 
order computation.By explicit computation we find that the sources listed in tables \ref{Rbasis} and \ref{Rbasisex}) '
are given at second order by the expressions we list in table \ref{Rbsrc2} below

\begin{center}
	\begin{table}[h!]
		\caption{Sources of $R_{MN}$ equations at 2nd order}\label{Rbsrc2}
		\resizebox{\columnwidth}{!}{
		\begin{tabular}{ |c| }
			\hline
			Scalar sector  \\ 
			\hline
			${\mS}^{S_1}(R) = e^{-R}(1-R)\left( \left( u \cdot K-u\cdot \nabla u \right)\cdot \CP \cdot \left( u\cdot K-u\cdot \nabla u \right) \right) $\\ 
			${\mS}^{S_2}(R)=  -\frac{1}{2} e^{-R} (R-2)\bigg(K_{MN}K_{PQ}P^{NP}P^{MQ}-\frac{\mathcal{K}^2}{D-3}\bigg) + \frac{1}{2} e^{-R}(R+2) \bigg(\nabla_Mu_N \nabla_Pu_Q P^{NP}P^{MQ}\bigg) $\\$ - \frac{e^{-R}}{2}\bigg(\nabla_{[M}u_{N]} \nabla_{[P}u_{Q]} P^{NP}P^{MQ}\bigg) -e^{-R} R\bigg(\nabla_Mu_N K_{PQ}P^{NP}P^{MQ}\bigg) $\\$ + \frac{1}{\mK}\frac{ e^{-R} (R-2) R}{4}\nabla^A\left( \frac{D-3}{\mathcal{K}} \left( \frac{D-3}{\mathcal{K}^3}\left( \nabla \nabla^2\mathcal{K}-\nabla^2\nabla^2u \right)+8(u\cdot K-u\cdot \nabla u)+u\cdot K+\frac{\nabla^2 u}{\mathcal{K}} \right)_{B} \CP_{A}^{B} \right) $\\$ - \frac{e^{-R} (R-2) R }{4}\frac{\nabla^2\nabla^2\mK}{\mK^3} + \frac{1}{4} e^{-2 R} \left(e^R \left(R^2+2 R-4\right)-2 (R-2) R\right)(u.\nabla u_M)(u.\nabla u_N)\CP^{MN} $\\$ + \frac{1}{2} e^{-2 R} \left(2 e^R (R-1)-(R-2) R\right) (u^AK_{AM})(u^BK_{BN})\CP^{MN} + e^{-2 R} (R-2) R (u.\nabla u_M)(u^CK_{CN})P^{MN} $\\$ + \frac{1}{4} e^{-R} (R-2) R \left(\frac{\nabla^2u_M}{\mK}\right)\left(\frac{\nabla^2u_N}{\mK}\right)\
CP^{
MN} -\frac{e^{-R} (R-2) R}{2}\left(\frac{\nabla^2u_M}{\mK}\right)(u.\nabla u_N)\CP^{MN}  $\\$ + \frac{1}{4} e^{-R} R \left(2 R^2-3 R-6\right)\frac{(u.\nabla\mK)^2}{\mK^2} - \frac{ e^{-R} \left(R^3-14 R^2+20 R+4\right)}{4}u.K.u\frac{\mK}{(D-3)}  $\\$ + \frac{ e^{-R} \left(3 R^3-38 R^2+62 R-4\right)}{4}\frac{\mK}{(D-3)}\frac{u.\nabla \mK}{\mK} - \frac{1}{4} e^{-R} R \left(R^2-6\right)u.K.u \frac{u.\nabla\mK}{\mK} + e^{-R} (R-1)\frac{\mK^2}{(D-3)^2}$\\$ -\frac{1}{4}e^{-R}~\left(\nabla_{(A}u_{B)}\nabla_{(C}u_{D)}\CP^{BC}\CP^{AD}\right) $\\
			${\mS}^{S_3}(R) = {\mathcal V}^{S_1}(R)-(1-e^{-R}){\mathcal S}^{S_2}(R) $ \\
			${\mS}^{S_4}(R) = (1-e^{-R}){\mathcal S}^{S_1}(R)- 2{\mathcal V}^{S_2}(R) $ \\
			\hline
			Vector sector  \\ 
			\hline
			 $ {\mS}^{V_1}_{M}(R) = \frac{1}{(1-e^{-R})}\left({\mathcal V}^{V}_L(R)-{\mathcal S}^{V_2}_L(R)\right) $ \\
			$  {\mS}^{V_2}_{A}(R) = \frac{\mK^2}{2(D-3)^2} \bigg[ -e^{-2 R} \left(e^R-1\right) \left(R^2-2\right)\frac{3}{2}\frac{D-3}{\mathcal{K}}\left( 1+2 \frac{u\cdot\nabla\mathcal{K}~(D-3)}{\mathcal{K}^2}-\frac{u\cdot K\cdot u~(D-3)}{\mathcal{K}}\right) \left( u\cdot \nabla u-u\cdot K \right)_{B}  $\\$ - e^{-2 R} \left(e^R-1\right) (R-1)\frac{(D-3)}{\mathcal{K}} \left( \frac{(D-3)}{\mathcal{K}^3}\left( \nabla \nabla^2\mathcal{K}-\nabla^2\nabla^2u \right)+8(u\cdot K-u\cdot \nabla u)+u\cdot K+\frac{\nabla^2 u}{\mathcal{K}} \right)_{B}  $\\$ + Re^{-R}\left( -2 \frac{(D-3)^2}{\mathcal{K}^2}\left( \frac{\nabla_{M}\mathcal{K}}{\mathcal{K}}-u^{D}K_{DM} \right)P^{MN} \left( \nabla_{N}u_{B}-K_{N B} \right)+\frac{(D-3)}{\mathcal{K}}\left( u^C K_{CB}-\frac{\nabla^2 u_B}{\mathcal{K}} \right) \right) \bigg] \CP_{A}^{B} $\\$ - \frac{e^{-R}}{2}\frac{\mK}{(D-3)}\bigg[ -{\mathcal E}_M +  D\frac{\nabla^2\nabla^2 u_M}{\mathcal{K}^3}-D\frac{\nabla_M (\nabla^2 \mathcal{K})}{\mathcal{K}^3} +3D\frac{(u\cdot K\cdot u)(u\cdot \nabla u_M)
}
{\mathcal{K}}-3D\frac{(u\cdot K
\cdot u)(u^A K_{AM})}{\mathcal{K}}$\\$-6D\frac{(u \cdot \nabla \mK)(u\cdot \nabla u)}{\mathcal{K}^2}+6D\frac{(u \cdot \nabla \mK)(u^A K_{AM})}{\mathcal{K}^2}+3 u \cdot \nabla u-3u^A K_{AM}\Bigg]\CP_L^M $\\
			\hline
			Tensor sector  \\ 
			\hline
			$ {\mS}^{T}_{LP}(R) = \bigg[  e^{-R} \frac{\mK}{(D-3)}  \left( (K_{MN}-\nabla_{(M}u_{N)})-\frac{\CP_{MN}}{D}(K_{AB}-\nabla_{(A}u_{B)})\CP^{AB} \right) $ \\$ -  e^{-R}  \left( (K_{MC}-\nabla_{C}u_{M})\CP^{CD}(K_{D N}-\nabla_{D}u_{N}) - \frac{\CP_{MN}}{D}(K_{AC}-\nabla_{C}u_{A})\CP^{CD}(K_{D B}-\nabla_{D}u_{B})\CP^{AB} \right) $\\$ -\frac{1}{2}e^{-2 R} \left(R^2-4 R+2 e^R (R-1)+2\right) \bigg( (u_{C}K^{C}_{M}- u.\nabla u_{M})(u_{C}K^{C}_{N}- u.\nabla u_{N})  $\\$ - \frac{\CP_{MN}}{D}(u_{C}K^{C}_{A}- u.\nabla u_{A})(u_{C}K^{C}_{B}- u.\nabla u_{B}) \CP^{AB} \bigg) \bigg]\CP^{M}_{L} \CP^{N}_{P} $ \\ 
			\hline
		\end{tabular}}
	\end{table}
\end{center}
\noindent
\newpage

 \begin{center}
	\begin{table}[h!]
		\caption{Sources to constraint equations at 2nd order}\label{Esrc2}
		\resizebox{\columnwidth}{!}{
		\begin{tabular}{ |c| }
			\hline
			Vector constraint source\\ 
			\hline
			$ {\mathcal V}_L^{V}(R) = \frac{1}{(D-3)} \nabla^P \bigg[ e^{-R}R \frac{D}{\mK} \bigg( (K_{MC}-\nabla_{C}u_{M})\CP^{CD}(K_{D N}-\nabla_{D}u_{N}) $\\$- \frac{\CP_{MN}}{(D-3)}(K_{AC}-\nabla_{C}u_{A})\CP^{CD}(K_{D B}-\nabla_{D}u_{B})\CP^{AB} \bigg)\CP^{M}_{L} \CP^{N}_{P} $\\$ - Re^{-R}\left( (K_{MN}-\nabla_{(M}u_{N)})-\frac{\CP_{MN}}{(D-3)}(K_{AB}-\nabla_{(A}u_{B)})\CP^{AB} \right)\CP^{M}_{L} \CP^{N}_{P}  $\\ $ + (e^{-2 R} \left(e^R-1\right) (R-2) R) \frac{(D-3)}{2\mK}  \bigg( (u_{C}K^{C}_{M}- u.\nabla u_{M})(u_{C}K^{C}_{N}- u.\nabla u_{N}) $\\$ - \frac{\CP_{MN}}{(D-3)}(u_{C}K^{C}_{A}- u.\nabla u_{A})(u_{C}K^{C}_{B}- u.\nabla u_{B}) \CP^{AB} \bigg)\CP^{M}_{L} \CP^{N}_{P} \bigg] $\\$  - \frac{e^{-R}}{2}\frac{\mK}{(D-3)}\bigg[-{\mathcal E}_M + D\frac{\nabla^2\nabla^2 u_M}{\mathcal{K}^3}-D\frac{\nabla_M (\nabla^2 \mathcal{K})}{\mathcal{K}^3} +3D\frac{(u\cdot K\cdot u)(u\cdot \nabla u_M)}{\mathcal{K}}-3D\frac{(u\cdot K\cdot u)(u^A K_{AM})}{\mathcal{K}}$\\$-6D\frac{(u \cdot \nabla \mK)(u\cdot \nabla u)}{\mathcal{
K}
^2}+6D\frac{(u \cdot \nabla \mK)(u^A K_{AM})}{\mathcal{K}^2}+3 u \cdot \nabla u-3u^A K_{AM}\Bigg]\CP_L^M $\\ 
			\hline
			Scalar constraint 1 source\\ 
			\hline
			$ {\mathcal V}^{S_1}(R)$\\$ = \frac{ \left( e^{-2 R} R \left(e^R \left(R^2-6\right)+3 (R+2)\right)\right)\mK}{6(D-3) }\nabla^M\left( \frac{3}{2}\frac{(D-3)}{\mathcal{K}}\left( 1+2 \frac{u\cdot\nabla\mathcal{K}~(D-3)}{\mathcal{K}^2}-\frac{u\cdot K\cdot u~(D-3)}{\mathcal{K}}\right) \left( u\cdot \nabla u-u\cdot K \right)_{B}\CP_{M}^{B} \right) $\\$+ \frac{ \left( e^{-2 R} \left(e^R (R-2)+2\right) R\right)}{4\mK } \nabla^M \left( \frac{(D-3)}{\mathcal{K}} \left( \frac{(D-3)}{\mathcal{K}^3}\left( \nabla \nabla^2\mathcal{K}-\nabla^2\nabla^2u \right)+8(u\cdot K-u\cdot \nabla u)+u\cdot K+\frac{\nabla^2 u}{\mathcal{K}} \right)_{B} \CP_{M}^{B} \right) $\\$+\frac{ \left(- e^{-R} R^2\right)}{4 \mK} \nabla^M \left( \left( -2 \frac{(D-3)^2}{\mathcal{K}^2}\left( \frac{\nabla_{M}\mathcal{K}}{\mathcal{K}}-u^{D}K_{DM} \right)\CP^{MN} \left( \nabla_{N}u_{B}-K_{N B} \right)+\frac{(D-3)}{\mathcal{K}}\left( u^C K_{CB}-\frac{\nabla^2 u_B}{\mathcal{K}} \right) \right)\CP_{M}^{B} \right)$\\$-\frac{1}{4}
e^{-R}~\left(\nabla_{(A}u_{B)}\nabla_{(C}u_{D)}\CP^{BC}\CP^{AD}\right) + \frac{Re^{-R}}{2}\nabla^M {\mathcal E}_M $\\
			\hline
			Scalar constraint 2 source\\ 
			\hline
			$ {\mathcal V}^{S_2}(R) =  -\frac{1}{2} e^{-R} (R-1)\bigg(K_{MN}K_{PQ}P^{NP}P^{MQ}-\frac{\mathcal{K}^2}{D-3}\bigg) + \frac{1}{2}e^{-R}(3+R)\bigg(\nabla_Mu_N \nabla_Pu_Q P^{NP}P^{MQ}\bigg) $\\$ + \frac{1}{2}\left(-e^{-R}\right)\bigg(\nabla_{[M}u_{N]} \nabla_{[P}u_{Q]} P^{NP}P^{MQ}\bigg) -e^{-R} (1+R)\bigg(\nabla_Mu_N K_{PQ}P^{NP}P^{MQ}\bigg) $\\$+ \frac{1}{\mK}\frac{\left( e^{-R} (R+2) R\right)}{4}\nabla^M\left[ \frac{(D-3)}{\mathcal{K}} \left( \frac{(D-3)}{\mathcal{K}^3}\left( \nabla \nabla^2\mathcal{K}-\nabla^2\nabla^2u \right)+8(u\cdot K-u\cdot \nabla u)+u\cdot K+\frac{\nabla^2 u}{\mathcal{K}} \right)_{B} \CP_{M}^{B} \right] $\\$- \frac{\left(e^{-R} R^2\right)}{4}\frac{\nabla^2\nabla^2\mK}{\mK^3} + \frac{1}{4} \left(e^{-2 R} R\left(2+R(e^R-1)\right)\right)(u.\nabla u_M)(u.\nabla u_N)\CP^{MN}  $\\$- \frac{1}{4} \left(e^{-2 R}R(R-2) \right) (u^AK_{AM})(u^DK_{DN})\CP^{MN} $\\$+\frac{1}{4}e^{-R}R^2 \frac{\nabla^2 u_M}{\mK}\frac{\nabla^2 u_N}{\mK}\CP^{MN} - \frac{ \left(e^{-R} R^2\right)}{2}\frac{\nabla^2 u_
M}{\mK}u.
\nabla u_N\CP^{MN} $\\$+ \frac{1}{2}\left(e^{-2 R} R(-2+4e^R+R)\right) (u.\nabla u_M)(u^CK_{CN})\CP^{MN} $\\$+ \frac{\left(e^{-R} R \left(2 R^2- R-12\right)\right)}{4}\frac{(u.\nabla\mK)^2}{\mK^2} - \frac{\left( e^{-R} \left(R^3-14 R^2-8R+2\right)\right)}{4}u.K.u\frac{\mK}{(D-3)} $\\$+ \frac{ \left( e^{-R}R \left(3 R^2-32 R-2\right)\right)}{4}\frac{u.\nabla\mK}{(D-3)} - \frac{ \left(e^{-R} R \left(R^2-2R-18\right)\right)}{4} \frac{u.\nabla\mK}{\mK}u.K.u + e^{-R}R \frac{\mK^2}{(D-3)^2}  $\\$ -\frac{1}{4}e^{-R}~\left(\nabla_{(A}u_{B)}\nabla_{(C}u_{D)}\CP^{BC}\CP^{AD}\right) $\\
			\hline
		\end{tabular}}
	\end{table}
\end{center}
\newpage

\bibliographystyle{JHEP}
\bibliography{ssbib}
\end{document}